\documentclass[11pt]{revtex4}

\usepackage{graphicx}
\usepackage{dcolumn}
\usepackage{bm}
\usepackage{xcolor}
\usepackage[normalem]{ ulem }
\usepackage{amsmath}
\usepackage{amssymb}

\definecolor{macri_color}{HTML}{FFD700}
\definecolor{alberto_color}{HTML}{75AADB}
\definecolor{espert_color}{HTML}{3C3C3C}
\definecolor{lavagna_color}{HTML}{2830B4}
\definecolor{izq_color}{HTML}{EE0000}
\definecolor{massa_color}{HTML}{3C3C3C}
\definecolor{stolbizer_color}{HTML}{EE4D8B}

\graphicspath{{images/}{plots/}}

\begin{document}
    \flushbottom

\title{Evolution of the political opinion landscape during electoral periods}

\author{Tom\'as Mussi Reyero$^{1}$}
\author{Mariano G. Beir\'o$^{1,2}$}
\email[]{mbeiro@fi.uba.ar}
\author{\\J.Ignacio Alvarez-Hamelin$^{1,2}$}
\author{Laura Hern\'{a}ndez$^{3}$}
\author{Dimitris Kotzinos$^{4}$}
\affiliation{$^1$Universidad de Buenos Aires. Facultad de Ingenier\'{i}a, Paseo Col\'on 850, C1063ACV, Argentina}
\affiliation{$^2$CONICET -- Universidad de Buenos Aires. INTECIN, Paseo Col\'on 850, C1063ACV, Argentina}
\affiliation{$^3$Laboratoire de Physique Th\'{e}orique et Mod\'{e}lisation, UMR-8089 CNRS, CY Cergy Paris Universit\'{e}, 2 Av. Adolphe Chauvin, 95302, Cedex, France}
\affiliation{$^4$ETIS UMR-8051 CY Cergy Paris Universit\'{e}, ENSEA, CNRS,  2 Av. Adolphe Chauvin, 95302, Cedex, France.}

\begin{abstract}

We present a study of the evolution  of the political landscape during the 2015 and 2019  presidential elections in Argentina,  based on the data obtained from the micro-blogging platform Twitter. We build a semantic network based on the hashtags used by all the users following at least one of the main  candidates. With this network we can  detect the topics that are discussed in the society. At a difference with most studies of opinion on social media, we do not choose the topics a priori,
 they naturally  emerge from the community structure of the  semantic network instead. We  assign to each user a dynamical topic vector which measures the evolution of her/his opinion in this space and allows us to  monitor the similarities and differences among groups of supporters of different candidates. Our results show that the method is able to detect the dynamics of formation of opinion on different topics and, in particular, it can capture the reshaping of the political opinion landscape which has led to the inversion of result between the two rounds of the 2015 election.

\end{abstract}

\maketitle

\section{Introduction}

 Our understanding of how opinion is formed and evolves in society has benefited since the last decade from the rapidly increasing amount of data diffused by on-line social networks. In this way,  large scale, cross-cultural studies that
 were difficult to perform based on  standard off-line surveys become feasible.

 In spite of the different biases that are known to affect studies based on on-line social networks, in terms of age, gender, residence location, social status, etc., the enormous amount of information they convey remains useful in particular to detect trends in the evolution of social opinion, at least restricted to the users of these platforms whose amount increases continuously. 
 Moreover nowadays traditional broadcasting media, like radio or television, diffuse information or opinions selected from on-line social networks thus coupling this large but biased set of users with the general population.

The micro-blogging platform Twitter has been widely used in order to study the evolution of social opinion on different topics~\cite{gaumont2018reconstruction,boutet2013s}, as well as the properties of the social interaction networks that result from the different functionalities  offered by the platform (mentions, retweets, follower-followee)~\cite{himelboim2013tweeting,barbera2015birds}.
The intrinsic properties of the  platform, like the small  size of posts (tweets limited to 280 characters), or its simplicity of usage, (a message can easily be re-transmitted, a mentioned user can be alerted, etc.) make it an excellent tool to study situations where the time scale for the  opinion's evolution is short as in electoral processes~\cite{ahmed20162014,gaumont2018reconstruction,caldarelli2014multi,tumasjan2010predicting} or in social protests~\cite{borge2011structural,alvarez2015sentiment, howard2011opening}.

The possibility to predict  opinion evolution  using Twitter, which is of particular interest during an electoral campaign, has been seriously  challenged~\cite{murthy2015twitter,gayo2012no,chung2011can}. Nevertheless, this kind of studies remain interesting because of their explanatory power.
For instance, it was possible to unveil the  spontaneous character of the emergence of off-line demonstrations during the Spanish social movement 15M by correlating the intensity of posts at a given location with the different gatherings  observed off-line~\cite{borge2011structural}.
Also it was possible to  identify  the origin of a denigrating campaign against one of the potential candidates  to the last French presidential election by studying the community structure of a network of retweets~\cite{gaumont2018reconstruction}.

Twitter users share text messages, images and videos, and  they may  associate their posts to a concept, by the means of \textit{hash-tags} (words beginning by the character ``$\string # $"), in order  to install a given concept to be discussed on the public ground.

Among a vast literature devoted to studies of social phenomena based on Twitter data, one can identify those studies that concentrate on the structure of the different  networks that can be defined  (follower-followee, retweets, mentions and answers), and those which try to infer the opinion dynamics, based on text mining and analysis. Both aspects, structure and content, are nevertheless  entangled, as users are mainly exposed to the content produced by those other users that they have decided to follow or by those selected by the algorithms of the platform~\cite{himelboim2013tweeting,barbera2015birds}.
This situation has been shown to lead to the  phenomenon  known as  \emph{echo chambers} or \emph{bubbles}, meaning that in fact, in the  worldwide open field of Twitter (as well as that of other internet platforms) users are mostly and sometimes uniquely exposed to the same information, frequently the one that comforts their own opinion, thus limiting the  possibilities of a real discussion~\cite{kelly2018vision,nikolov2015measuring,eady2019many}.

An interesting way to study the structure of opinions in Twitter exploits the hashtags chosen by the users,  assuming that this choice reveals a concept that the user wishes to address.  In a recent work~\cite{cardodo2019topical}, topics are defined by determining the community structure in a weighted network of hashtags, where two hashtags are connected if they appear together in the same tweet.
Assuming that the coexistence of hashtags is semantically meaningful, the community structure of such network  can reveal the general topics under discussion.
In this way,  the users may be characterized by a topic vector, with a dimension equal to the number of communities detected and  where each component informs about the interest of the user on  the different topics. The authors show that the similarity among users connected by a follower-followee relationship or by a mention relationship is higher on average  than the similarity among a sample of random users.

In this work we  extend the ideas developed in~\cite{cardodo2019topical} to a dynamical study of the  rapidly evolving opinion landscape that takes place in a society during an electoral campaign. This method allows us to recover the dynamics of the political tendency without introducing questions to the population, which are known to be subject to different bias (of formulation, false declarations, etc.) and {\em without imposing a priori}, neither an ontology nor the number of topics to be inspected. Our method just extracts the information coded in the data with the only assumption that two hashtags used in the same tweet are semantically related.
Our results show that, in spite of the limitations of  studies of opinion using Twitter described above, this method is able to capture the opinion evolution of the users with a  high enough time-scale resolution so as to  detect, for example,  the  reconfiguration of the political landscape taking place in the short period between the first and second round of the election, which, in one of the cases presented here, overturned the score of the first round of the election.

\section{Methods and Data set}
\label{data}

\subsection{Data Capture}
\label{data_capture}

This study is based on data captured during the two recent Argentinian presidential campaigns, in 2015 and in 2019. The periods of data capture extend from July 1st 2015 to March 31st 2016, and from January 1st, 2019 to December 10th 2019, the main elections being held on October 25th, 2015 and October 27th, 2019 (see the SI for a  detailed description of the electoral processes).

The capture is based on the \emph{active}  followers (we define a user as active in Twitter if he/she posted at least one tweet during the first month of capture)
of the candidates for president or for deputy-president of each of the main political parties participating in these  elections.
We filter those users whose profile location is set to some city/province in Argentina, in order to focus on those Twitter users that are residents in the country, and we capture and process all their tweets in the period. Table~\ref{table_statistics} gives a summary of the basic statistics for each dataset.

A detailed  description is provided in the SI regarding the number of tweets captured daily and their classification as  original tweets, simple retweets, retweets with comment and replies along with an  analysis of the geographical and gender distribution of our user base.

\begin{table}
\centering
\caption{Basic statistics for each electoral process.}
\label{table_statistics}
\begin{tabular}{lll}
\hline\noalign{\smallskip}
 & 2015 elections & 2019 elections  \\
\noalign{\smallskip}\hline\noalign{\smallskip}
Number of active argentinian users (AAU) & 218k &586k \\
Number of tweets by AAUs captured & 50M & 280M\\
Time frame & Jul. 1st, 2015 to Mar. 31st, 2016 & Jan. 1st, 2019 to Dec. 10th, 2019\\
Primaries & August 9th, 2015 & August 11th, 2019 \\
First round & October 25th, 2015 & October 27th, 2019 \\
Second round (ballotage) & November 22nd, 2015 & -- \\
\noalign{\smallskip}\hline
\end{tabular}
\end{table}

\subsection{Definition of topics  and user's description vectors}
\label{topics_definition}

Hashtags are keywords created and chosen by the users, which can be interpreted as representing the engagement of users with events, ideas or different discussion subjects. If two hashtags highly co-occur (i.e., they  frequently appear together in the same tweet) it  is a reasonable hypothesis to assume a semantic association between them.
Following the ideas developed in~\cite{cardodo2019topical}, we build a complex weighted network based on hashtags' co-occurrence. Then, the topics of discussion arise as communities measured on this network, which we detect using the OSLOM algorithm~\cite{lancichinetti2011finding}. It is worthwhile noticing that the topics simply \emph{emerge} from the community detection algorithm which is completely agnostic regarding their meaning and does not pre-determine their number.

We describe the interests of each user $i$
 by means of a user description vector $\boldsymbol{d_i}$ of dimension $N_T$,
the number of topics (communities) found, which informs about the topic preferences of  user $i$.

This description vector is computed in the following way:
\begin{enumerate}
    \item  We build a user-topic matrix, $U$, where each element, $u_{ij}$, gives the absolute number of times that user $i$ has used a hashtag that belongs to the community identified as  topic $j$.
    \item  We compute the global  topic vector $\boldsymbol{T}=\sum_i ^{N}{\boldsymbol{u_i}}$, where $\boldsymbol{u_i}$ is the $i$-th row vector in the user-topic matrix, and $N$ the size of the population. This vector gives the total number of times that each topic has been used by all the users in the dataset.
    \item  We define the vector $\boldsymbol{v_i}$ which gives  the difference between the frequency of usage of the topic by user $i$ and its global  frequency of usage in the population.

\begin{equation}
   \boldsymbol{v_i}  = \frac{\boldsymbol{u_i}}{||\boldsymbol{u_i}||_1} - \frac{\boldsymbol{T}}{||\boldsymbol {T}||_1} \enspace.
   \label{frequency_vector}
\end{equation}

Here the norm $||.||_1$ must be understood as the sum over all the components in the space of dimension $N_T$. The vectors of Eq.~\ref{frequency_vector} thus inform about whether user $i$ has addressed each of the identified topics more or less than on average.
\item As we are only interested in the orientation of the description vectors, they are normalized as:
\begin{equation}
    \boldsymbol{d_i} = \frac{\boldsymbol{v_i}}{||\boldsymbol{v_i}||_2} \enspace,
\end{equation}

where $||\boldsymbol{v_i}||_2$ is the standard euclidean norm in the topic hyperspace of dimension $N_T$.


\end{enumerate}


 \paragraph{Dynamical measurements.} In order to track the evolution of the users' interests we apply the aforementioned procedure to sliding time windows of 7 days, thus producing a series of matrices $U_t$, one for each day. We shall call $\boldsymbol{d}_i^t$  the description vector for user $i$ at discrete time $t$.

\begin{figure*}[h!]
  \centering
  \includegraphics[height=9.6cm]{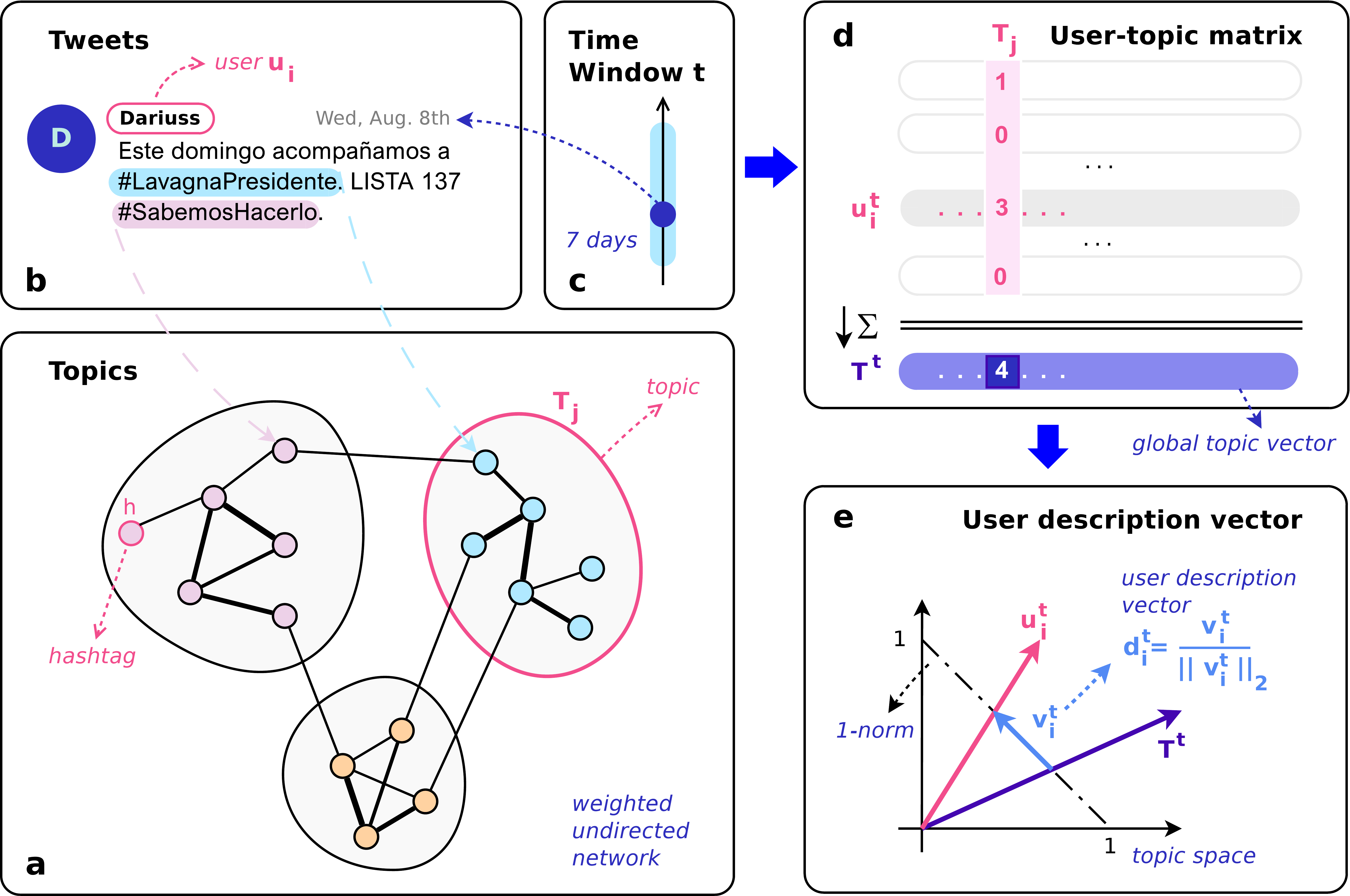}
 \caption{\label{procedure_illust}\textbf{Illustration of the procedure used to  compute the dynamical user's description vectors.}
 The semantic network (panel a) is built by connecting two hashtags that appear together in the same tweet (panel b). The community structure of this  network determines the topics that are discussed in the platform (the colors of the nodes code the communities).  In order to obtain the dynamical user-topic matrices (panel d) we consider all the tweets in a sliding time window of seven days (panel c); each row corresponds to a single user and codes in the columns the number of times that the user has used one hashtag belonging to the corresponding topic during the considered period.
  The normalized sum over all the users of each column of the user-topic matrix (panel d)  gives the average usage of each topic as a function of time, while the rows of the matrix give all the topics discussed by a single user. Finally, for each user one can obtain the vector $\Vec{d_i^t}$ (panel e) which gives the difference between the topics interesting user $i$ at time $t$ and the average usage of the topics over the population.
 }

\label{diagram}
\end{figure*}

The full procedure is illustrated in Fig.~\ref{procedure_illust}.

\subsection{Measuring the similarity between groups of users}

We define the similarity between a pair of users $i$ and $j$ as  the cosine similarity between the corresponding description vectors. As the latter are normalized, the similarity reduces to the inner product:
\begin{equation}
    s(i,j) = \langle \boldsymbol{d_i} , \boldsymbol{d_j} \rangle \enspace.
\end{equation}

We also define the average description vector of a group of users $G$,  of cardinal $|G|$:

\begin{equation}
   \boldsymbol{D_G}  = \frac{\sum_{i \in G} \boldsymbol{d_i}}{|G|} \enspace.
\end{equation}

Now we can introduce two indices measuring \textit{collective similarities}:

\begin{itemize}
    \item The \textit{cohesion} of a group of users, \textit{intra-group similarity} or \textit{self-similarity}, $s(G,G)$, defined as the average similarity between all its users, and computed in the following way:

\begin{equation}
    s(G,G) =
       \frac{\sum_{i,j \in G}
         {s(i,j)}}{|G|^2} =
    \frac{\sum_{i \in G}{\langle \boldsymbol{d_i}, \boldsymbol{D_G}  \rangle}}{|G|}={\langle \boldsymbol{D_G}, \boldsymbol{D_G} \rangle}=||\boldsymbol{D_G}||^2 \enspace.
\end{equation}

    \item The \textit{cross-group similarity} is the average similarity between  members of different groups $G_1$ and $G_2$, namely $s(G_1, G_2)$:

\begin{equation}
    s(G_1,G_2) =
       \frac{\sum_{i\in G_1,j \in G_2}
         {s(i,j)}}{|G_1|\cdot|G_2|} =
    \frac{\sum_{i \in G_1}{\langle \boldsymbol{d_i}, \boldsymbol{D_{G_2}} \rangle}}{|G_1|}={\langle \boldsymbol{D_{G_1}}, \boldsymbol{D_{G_2}} \rangle} \enspace.
\end{equation}

\end{itemize}

\section{Results}

The results presented in this section are based on  tweets collected as described in Section~\ref{data}, for the two last presidential elections in Argentina (details of the retrieval and cleaning methods  of the data-set can be found in the SI).

As explained in section~\ref{topics_definition},   we built the semantic network with the  assumption that hashtags used in the same tweet carry some semantic similarity. The community structure of such network reveals the topics that are discussed in the society, and the description vectors allow us to characterize the interests of each user, in the topic space.

\begin{figure*}[h!]
  \centering
  \includegraphics[height=6.4cm]{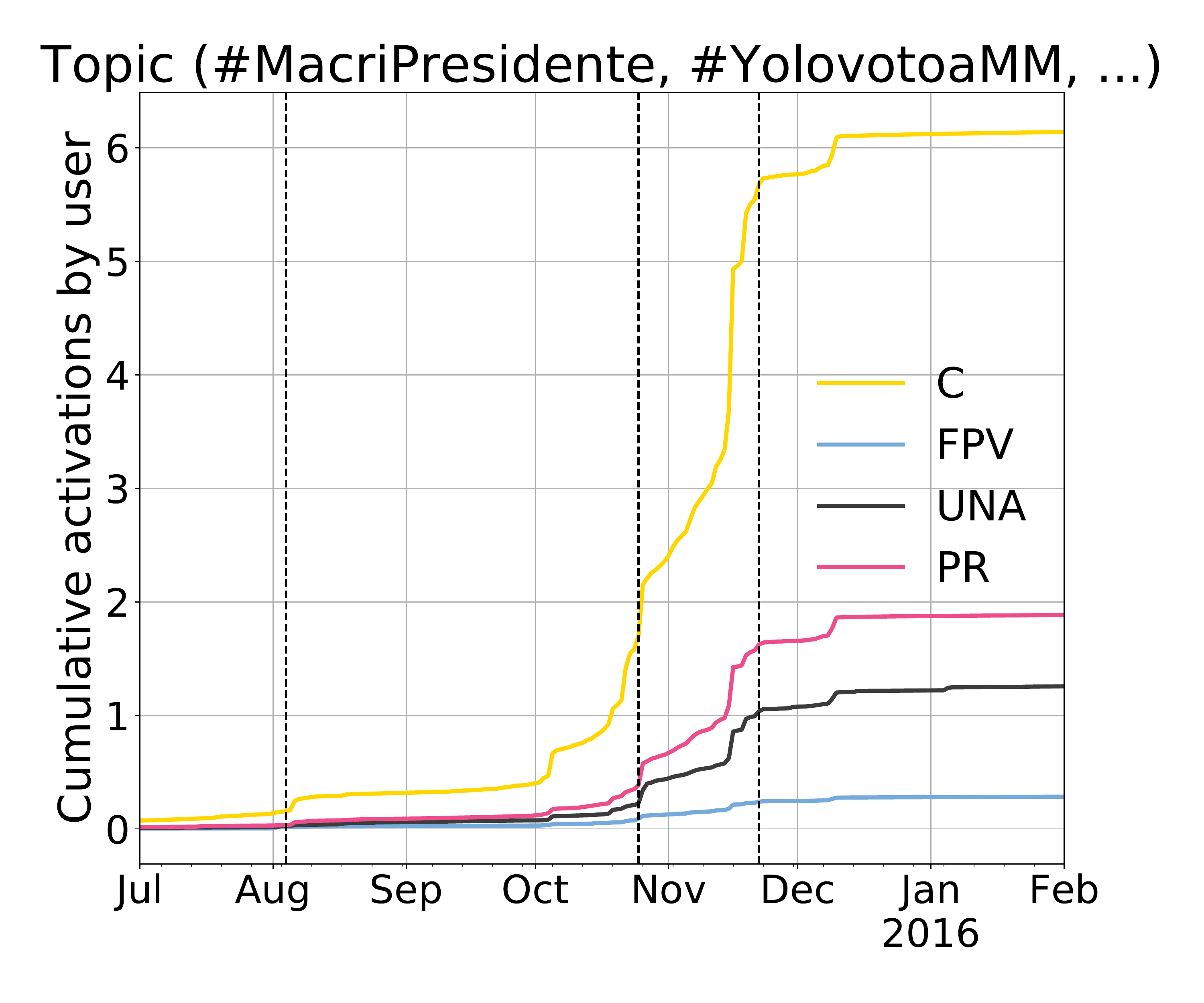}
  \includegraphics[height=6.4cm]{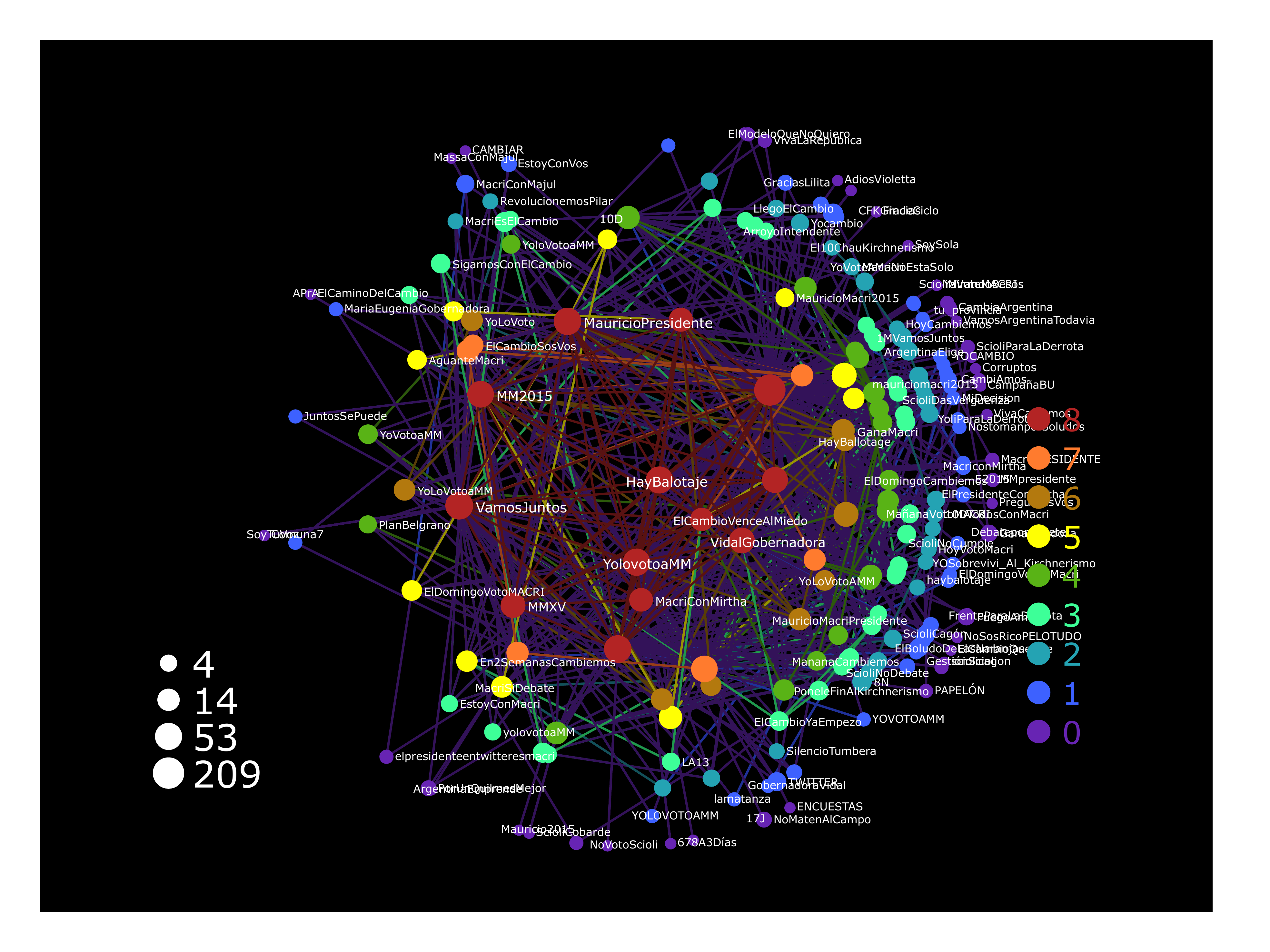}
\caption{\textbf{Main topic  supporting  the C party during the second round of the 2015 elections.} Left panel:
Cumulative usage of the topic by the supporters of the different parties. Right panel:   Hashtag sub-network of the topic. Nodes are arranged according to the $k$-core decomposition of the community graph. Figure produced with LaNet-vi ~\cite{lanet-vi}.}
\label{topic_63_evolution2015b}
\end{figure*}

In Fig.~\ref{topic_63_evolution2015b} we show the structure of a topic (right panel) composed of hashtags supporting the \textit{Cambiemos} party (C), one of the two major parties in the second round of the 2015 election. As the topics emerge from the community analysis without any a priori information introduced into the system (they are arbitrarily labelled by a number), it is the inspection of the hashtags included in each community that informs about the subject to which each topic is related.  On the left panel we show the cumulative number of supporters of each political party referring to that topic. This shows that, although it is not always true  that people choose a hashtag only to support the idea it conveys  (notice that members of other parties, including the strongest opponent, FV,  also use the considered topic), on average,  our method does correctly capture the expected preferences of the users. On the right panel, the $k$-core structure of the sub-network of hashtags that compose this topic is shown. The $k$-core decomposition represents a graph as a series of layers (the cores) in which the core of index $k$ is the maximal induced subgraph with minimum degree $k$~\cite{alvarez2006large}. In our figures, the node color represents the coreness of each hashtag (the largest core to which it belongs), and its size represents its degree (the number of hashtags it is connected to).
A similar description of a topic in support of the largest opponent  party, FV, can be found in the SI.

Argentinian law imposes to the citizens the obligation to vote. As a consequence, not only high participation rates are observed but also political discussions occupy a significant part  of the public attention. In both elections the political opinion was highly polarized, with two main antagonist parties dominating the political spectrum. Other two or three smaller parties may, on certain occasions, play the role of a pivot for the determination of the final result; therefore understanding the evolution of the opinion of their supporters is a crucial issue.  We will see that this was the case of 2015 election, where a second round was necessary  to determine the winner. A second round  only takes place if  no party  obtains either \textit{(i)} more than 45\% of the votes or; \textit{(ii)} more than 40\% and a 10\% difference with the second most voted party.

 Unlike in  2015 election, no second round took place in 2019.  In fact that year, the primary elections played the role of a first round. The electoral rule establishes that the primary elections, called PASO, \emph{Primarias Abiertas, Simultaneas y Obligatorias} meaning ``open, simultaneous, and compulsory primary elections'', take    place simultaneously and all the competing parties are bound to present at least one candidate. Due to the particular  political configuration of that moment, no party took the  risk to divide its votes into several candidates, and therefore they  presented one single candidate each. Under such circumstances, these primaries were seen as rehearsal of the general election.

 Both situations are excellent case studies for this work because they present  a short time window with a fast dynamics of  political opinion, allowing us to detect the reconfiguration of the opinion landscape at two different elections, corresponding to different political situations, thus confirming the validity of  our method.

 \subsection{ 2019 Elections}


The two main parties intervening in  this election were \emph{Juntos por el Cambio} (JPC), whose candidate was the incumbent,
and the challenger (and previous ruler) \emph{Frente de Todos} (FDT).

Table~\ref{parties_2019} describes the main participant parties along with their acronyms and a rough characterization of their position in the political spectrum.

\begin{table}[h]
    \centering
    \begin{tabular}{c|l|c|l|c|c}
       & {\bf Party Name}   & \bf Acronym  & \bf Political/economical orientation & \bf PASO & \bf 1st Round \\
       \hline
       \textcolor{macri_color}{$\blacksquare$} & Juntos por el Cambio (in power)
       &  JPC  & Center-right, economically liberal  & $31.80\%$& $38.67\%$   \\

       \textcolor{alberto_color}{$\blacksquare$} & Frente de Todos (challenger)   &  FDT  & Center-left, economically interventionist  & $47.79\%$ & $48.24\%$  \\

       \textcolor{lavagna_color}{$\blacksquare$} & Consenso Federal    &  CF  & Center, economically liberal & $8.15\%$ & $6.14\%$  \\

       \textcolor{izq_color}{$\blacksquare$} & Frente de Izquierda   &  FI  & Leftist party & $2.83\%$ & $2.16\%$   \\

       \textcolor{espert_color}{$\blacksquare$} & Frente Despertar     &  FD  & Right party          & $2.16\%$ & $1.47\%$  \\
    \end{tabular}
    \caption{Acronyms and characteristics of the main contenders of 2019 presidential election.}
    \label{parties_2019}
\end{table}{}

\begin{figure*}
  \centering
  \includegraphics[width=16cm]{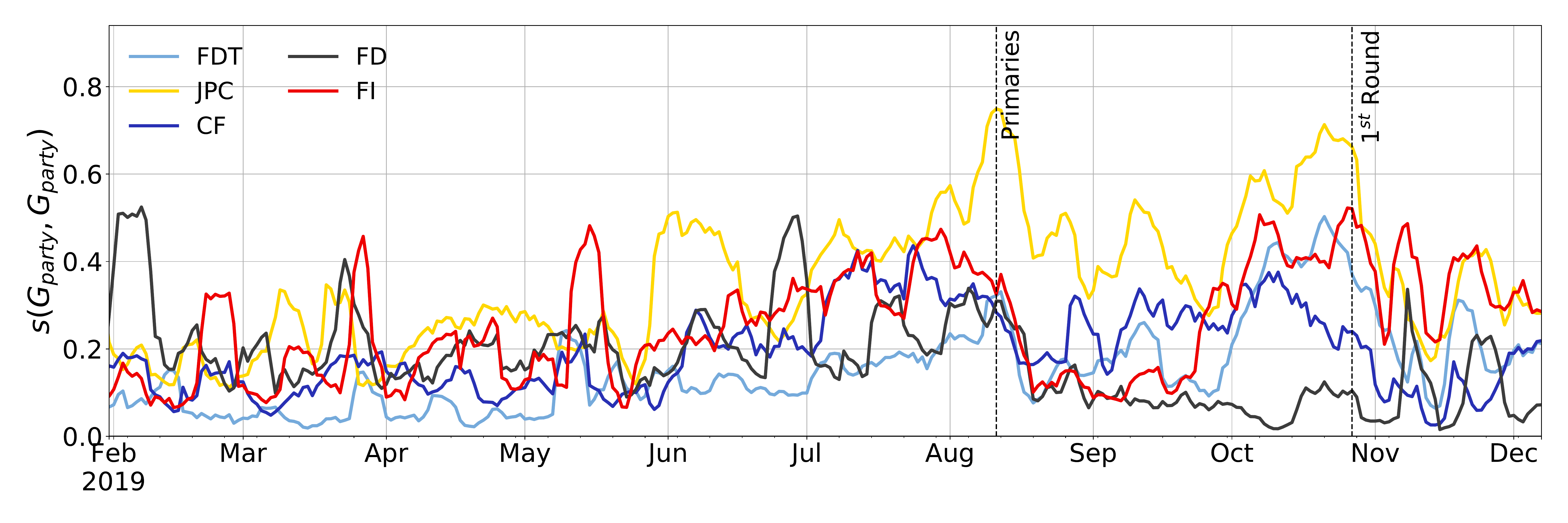}
  \caption{\textbf{Self-similarity of the supporters of the main parties listed in table~\ref{parties_2019}.}}
  \label{self_similarity_2019}
\end{figure*}

Fig~\ref{self_similarity_2019} shows the self-similarities for the parties participating to the 2019 election  as a function of time.

The ruling party shows, in general,    a higher self-similarity than the others which increases as the PASO approaches, reflecting the strong cohesion of its supporters. This may be interpreted as the strong need to defend their governmental choices face to its several opponents. Interestingly,  peaks of strong self-similarity are observed at different times,  for the two minority parties at both extremes of the political spectrum,  FI and FD. This is the signature of the occurrence of a particular event which  triggers a coherent reaction among the supporters of one party while for the supporters of the other parties the reaction to that  event is not discriminating. This happens when the event is in resonance with the political traditions of one of these  parties.

It is worthwhile noticing that the description vectors of the users contain a large diversity of topics,  many of which do not have a political character. When the public discussion is dominated by one of these topics  (for instance a football championship) the differences among the supporters of different parties may be partially and temporarily washed out. This effect is enhanced far from the electoral dates, where we can observe that the all the parties fluctuate around the same value of self-similarity, except for the isolated peaks already mentioned.

In order to further investigate the observed isolated  peaks, it is necessary to proceed to  a careful inspection of the dominant topics at the considered date. This can be done using the platform we have created to analyze the evolution of the different topics~\cite{platform}.  Let us consider, as an example, the two sharp peaks observed in Fig.~\ref{self_similarity_2019}, by the end of March 2019 which correspond to the curves of FD and FI (black and red respectively).

\begin{figure*}[h!]
  \centering
  \includegraphics[height=5.9cm]{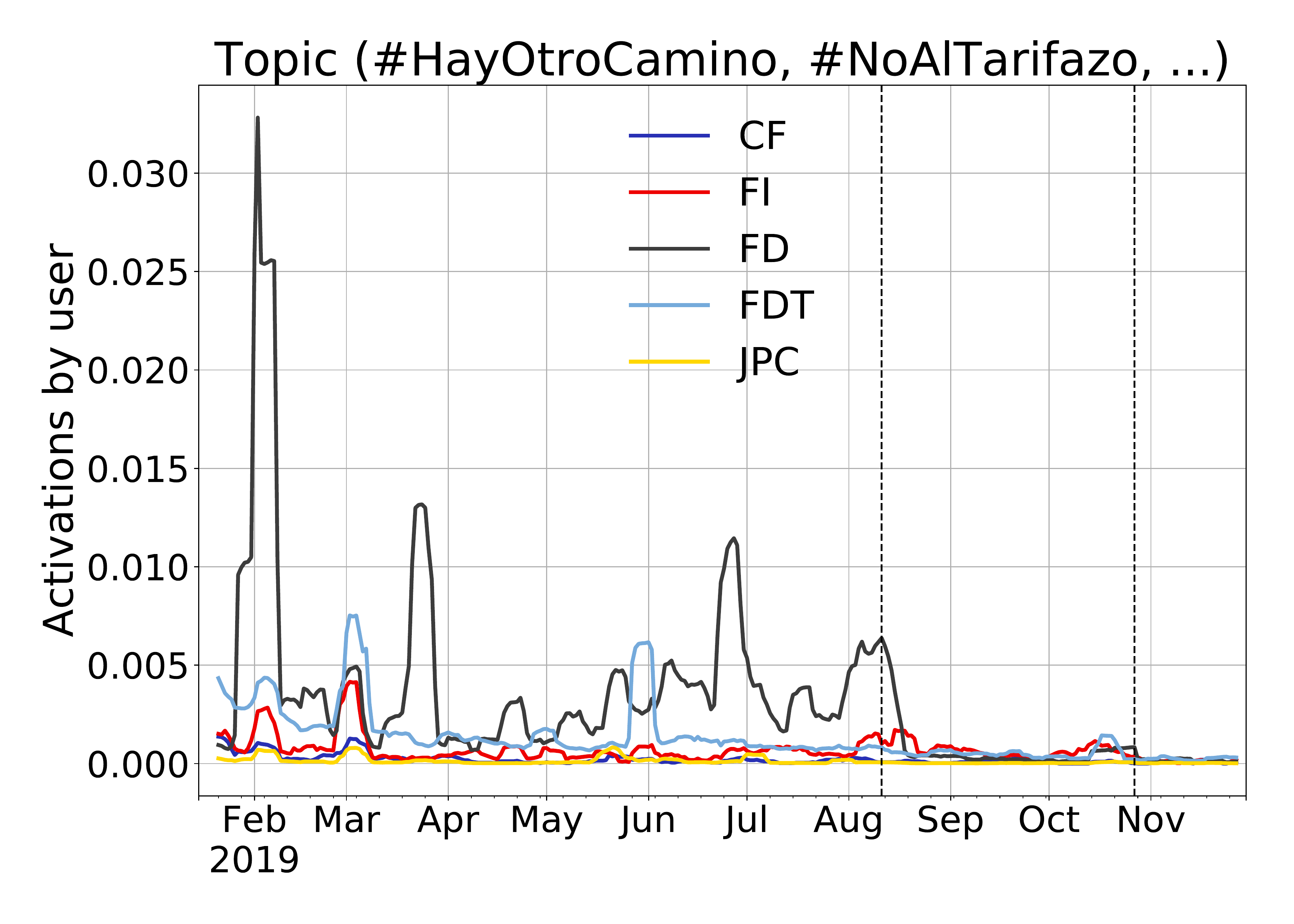}
  \includegraphics[height=5.9cm]{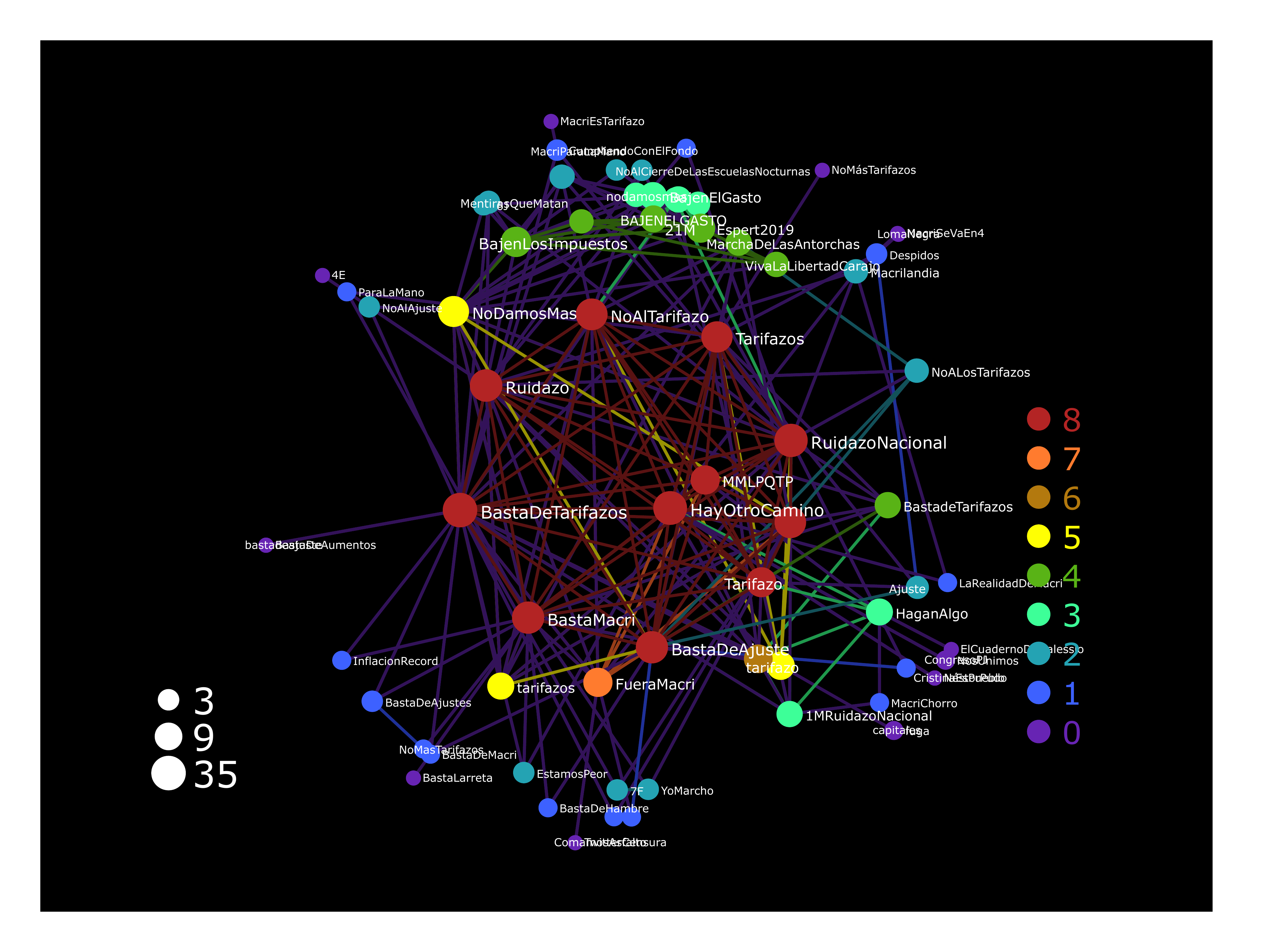}
\caption{\textbf{Discussions during 2019 electoral period. Topic complaining of high taxes while JPC party was ruling.} The vertical dotted lines indicate the dates of the primary and main elections. 
Left: Time evolution of the topic usage by the supporters of the different parties (7-day rolling average). 
Right: Hashtag composition of the topic. Nodes are arranged according to the $k$-core decomposition of the community graph. Figure produced with LaNet-vi ~\cite{lanet-vi}.}
\label{topic_69_evolution2019}
\end{figure*}

Fig.~\ref{topic_69_evolution2019} shows that the   peak of the FD self-similarity curve, located at March 21st 2019 (in black) corresponds to a demonstration against tax rises which took place that day in front of the National Congress.
This is a subject that usually interests right and liberal parties, although JPC supporters (mainly liberals) did not comment on it on the same terms,  because their party being in power, was  responsible for the tax rise. In fact,  an inspection of the topics that interested JPC supporters at that time using the platform~\cite{platform} (cf. the smaller peak of the JPC  curve  (yellow)   observed  in Fig.~\ref{self_similarity_2019}), shows that this small peak does not correspond to the topic analysed here.

\begin{figure*}[h!]
  \centering
  \includegraphics[height=5.9cm]{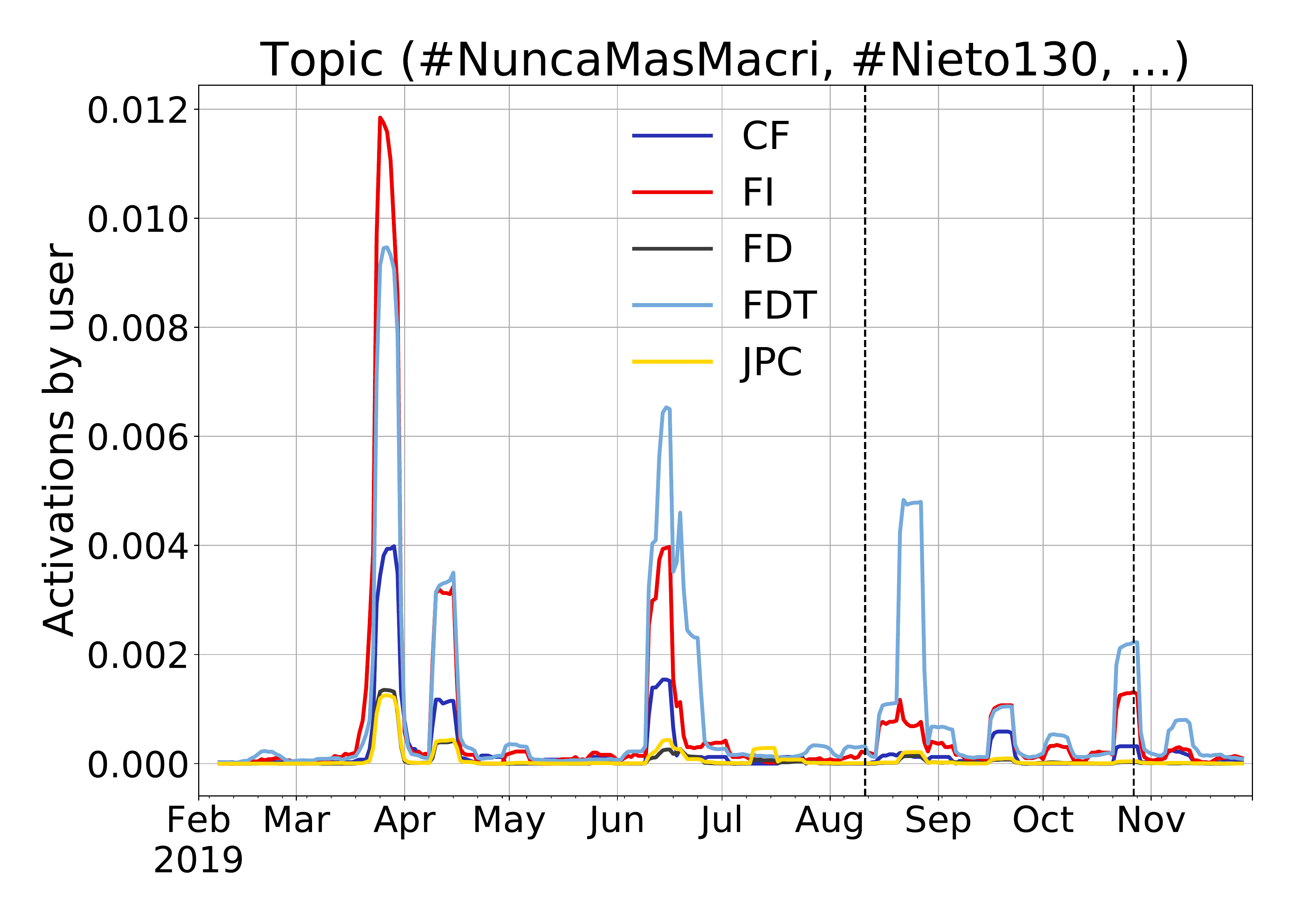}
  \includegraphics[height=5.9cm]{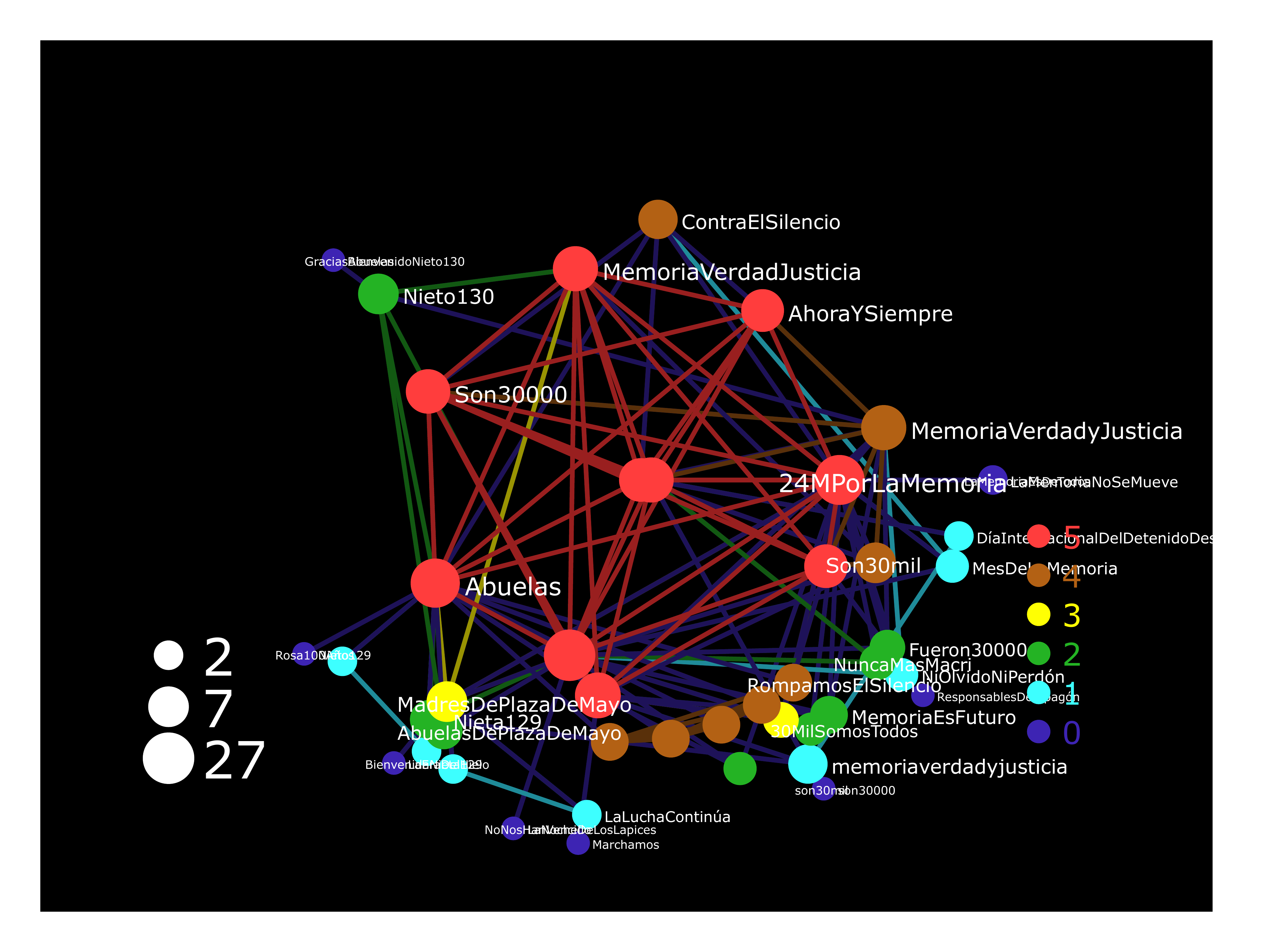}
\caption{\textbf{Discussion during 2019 electoral period. Topic evoking the anniversary (March 24th, referred as M24) of the installation of the last dictatorship.}  The vertical dotted lines indicate the dates of the primary and main elections. 
Left: Time evolution of the topic usage by supporters of the different parties (7-day rolling average). 
Right: Hashtag composition of the topic. Nodes are arranged according to the $k$-core decomposition of the community graph. Figures produced with LaNet-vi ~\cite{lanet-vi}.}
\label{topic_67_evolution2019}
\end{figure*}

Fig.~\ref{topic_67_evolution2019} depicts the temporal behaviour of the topic involved in
 the other peak of Fig.~\ref{self_similarity_2019}, situated around  March 24th (FI curve, in red). This day celebrates the remembrance of the victims of the last dictatorship in Argentina (1976-1983). The leftist parties (FI and FDT) are the most concerned with the topic during that day, while the more conservative JPC and FD (right wing) are scarcely active in the topic. Interestingly this topic is later reactivated periodically, but mainly due to the activity of FDT, which is a major, composite party including an important left wing.

So, the inspection of the topics shows that the three peaks observed  very near the end of March 2019 in the self-similarity curve correspond  to different discussions. In this way the evolution of the  self similarity captures the dynamics of the important topics discussed on the platform. Other significant peaks have been equally identified and the interested reader can find a more detailed description in the SI, as well as a dynamical inspection of the topics in the platform for topic visualisation and analysis that we have created~\cite{platform}.

\begin{figure*}
  \centering
    \includegraphics[width=16cm]{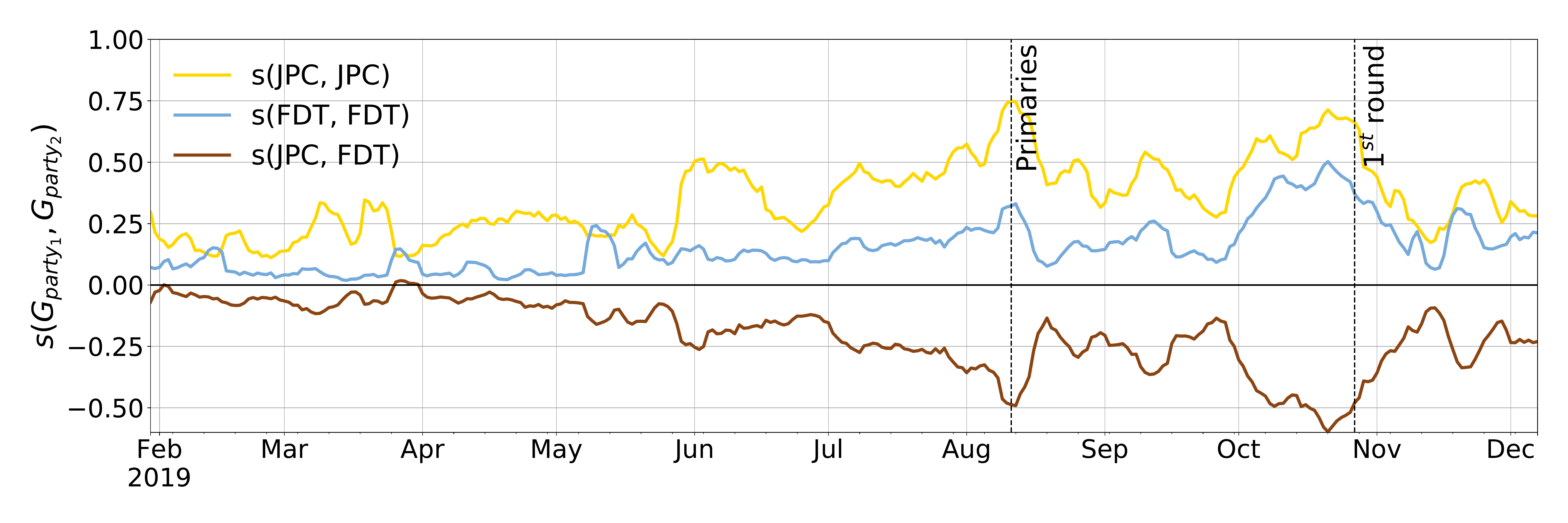}
  \caption{\textbf{ Cross-similarity between the supporters of the two main competing parties in 2019}. The vertical dotted lines indicate the dates of the primary and main elections. In brown, temporal evolution of the cross-similarity between the supporters of the JPC and FDT parties. The self-similarities of the two parties are also shown for the sake of comparison 
 (yellow and light blue curves).}
\label{cross_sim_main2019}
\end{figure*}

Cross-similarity curves  reveal other aspects of the evolution  of the opinion during the electoral process.
The brown curve in Fig.~\ref{cross_sim_main2019} shows the cross similarities between the two main antagonistic parties in 2019, JPC and FDT, compared to  the respective self similarities (i.e., the same curves shown in Fig.~\ref{self_similarity_2019}), plotted as a reference. As expected, we observe that the higher the self-similarities, the lower the cross-party one.
The cross-similarity strongly decreases in the vicinity of  the primary and the main elections.

\begin{figure*}[h!]
  \centering
  \includegraphics[width=16cm]{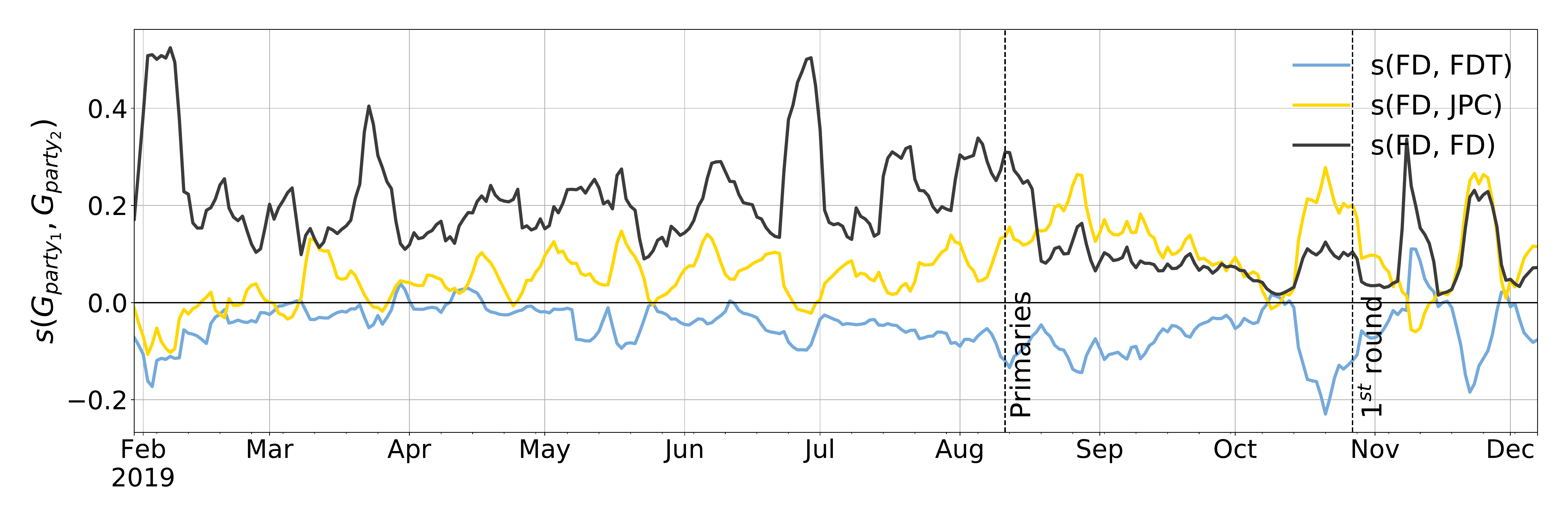}
  \includegraphics[width=16cm]{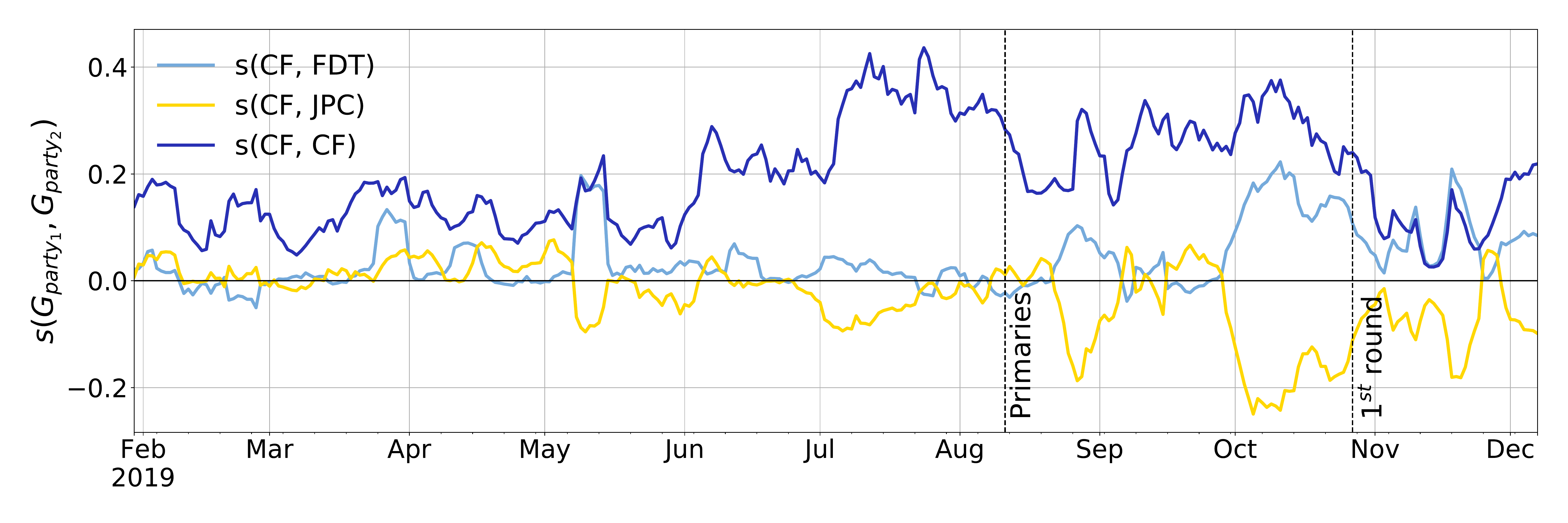}
  \includegraphics[width=16cm]{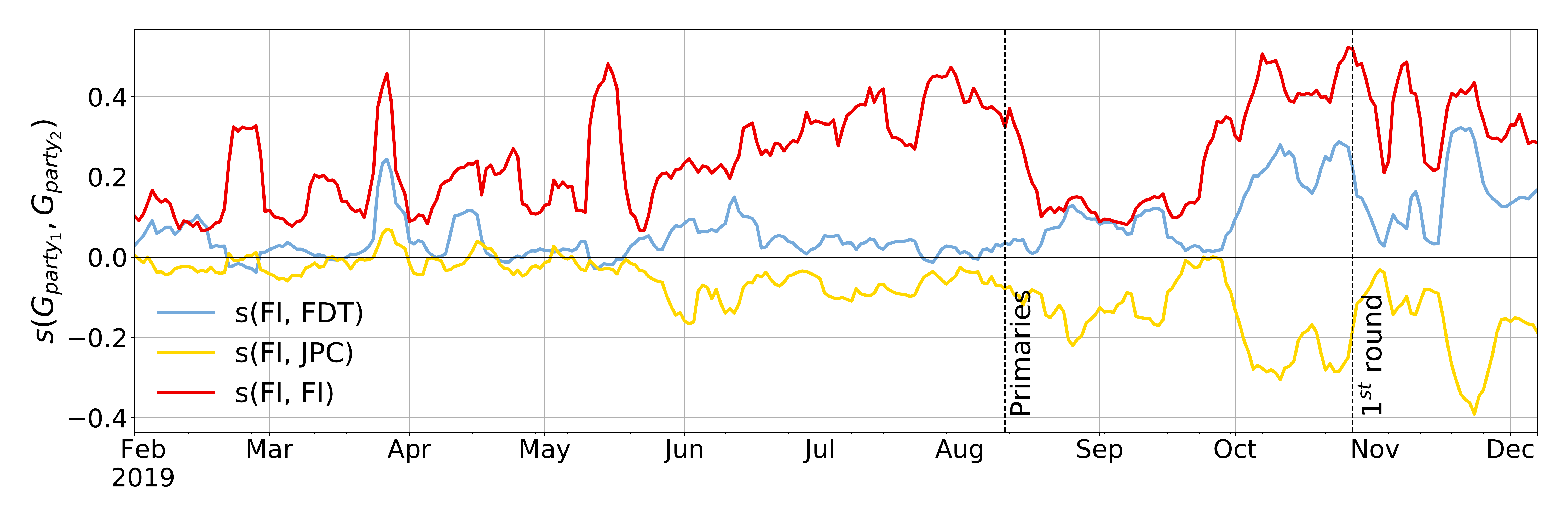}
\caption{\textbf{ Cross-similarity between the supporters of each of the minority parties and those of the two major ones in 2019,  JPC (in yellow) and FDT (in light blue)}. In each panel the self-similarity of the corresponding minority party is reproduced as a reference. 
Top: Cross-similarity of FD against
 the two main parties. 
 Center: Cross-similarity of CF against the two main parties. 
 Bottom: Cross-similarity of FI against the two main parties.}
\label{cross_sim_min_maj2019}
\end{figure*}

The inspection of the  dynamical  behaviour of the cross-similarities between  supporters of the  minority parties with the two major ones is particularly interesting, as it reveals to  what extent they contribute to   influence the final outcome.
Fig.~\ref{cross_sim_min_maj2019} shows a clear difference of behaviour among smaller  parties. In the top panel, while FD (Right party) has, most of the time and in particular near the electoral periods, a positive cross-similarity with the JPC in power, it holds almost always a negative cross-similarity with the challenger (FDT). Interestingly by mid-november 2019 this tendency is reverted, and a (small) positive similarity with FDT is observed at the same point where a strong self-similarity appears for the supporters of FD party. An inspection of the active topics of the moment reveals that this activity is related with Latin-american international politics, which mobilizes both parties but on opposite sides of the opinion spectrum. This shows that although we do  not perform any sentiment analysis here, the attention of the groups about a given subject is correctly captured independently of the opinion of the users on that subject.


On the contrary, the supporters of CF (dissident liberal party) show a complete different behaviour. Both cross-similarities with the two major parties are  quite low and fluctuating far from electoral periods. This tendency remains before the primary elections showing the strong support of their own candidate. However, before the first round a strong positive similarity with FDT develops (after some fluctuating period), showing that CF supporters are ready to support  the challenger FDT, from the first round.  

Finally Fig.~\ref{cross_sim_min_maj2019} shows the cross-similarity between the FI party and the two main competitors. Although  the leftist party has a very small influence in the country, the interest of this curve is to reveal that indeed its cross-similarity with FDT, is clearly positive (and negative with JPC). This is a clear evidence of the existence of a strong leftist component in FDT, as  mentioned above, which is completely absent in JPC.

 \subsection{Elections 2015}

The discussion of the results concerning this previous election, is  interesting not only as a test of  the validity of   our method in a different political context, but also because unlike in 2019,  this election required two rounds  to establish the winner. Even more interesting,  the rank  of the two first qualified parties was overturned in the second round.
We will show that our method is able to identify how the details of the dynamics of the opinion of the supporters of the smaller parties contributed to this final result.

 The main  political parties intervening in this election, which are formally different from those of the 2019 election, although there are important overlaps,  are listed in Tab.~\ref{parties_2015}.
\begin{table}[h]
    \centering
    \begin{tabular}{c|l|c|l|c|c|c}
       & {\bf Party Name}   & \bf Acron.  & \bf Political/econ. orientation & \bf PASO & \bf 1st Round & \bf 2nd Round \\
       \hline
       \textcolor{alberto_color}{$\blacksquare$} & Frente para la Victoria (in power)  &  FPV  & Center-left, econ.interventionist  & $38.67\%$ & $37.08\%$ & $48.66\%$ \\

       \textcolor{macri_color}{$\blacksquare$} & Cambiemos (challenger)
       &  C  & Center-right, econ. liberal  & $30.12\%$ & $34.15\%$ & $51.34\%$  \\

       \textcolor{massa_color}{$\blacksquare$} & Unidos por una Nueva Alternativa    &  UNA  & Economically interventionist & $20.57\%$ & $21.39\%$ & - \\

       \textcolor{stolbizer_color}{$\blacksquare$} & Progresistas   &  PR  & Socially and econ. liberal & $3.47\%$ & $2.51\%$ & -  \\

    \end{tabular}
    \caption{Acronyms and characteristics of the main contenders of 2015 presidential election.}
    \label{parties_2015}
\end{table}{}

The dynamics of the self-similarities, as well as that of the cross-similarity between the two largest parties,  follow a similar pattern as those of 2019 elections and are detailed in the SI.

\begin{figure*}[h!]
  \centering

  \includegraphics[width=16cm]{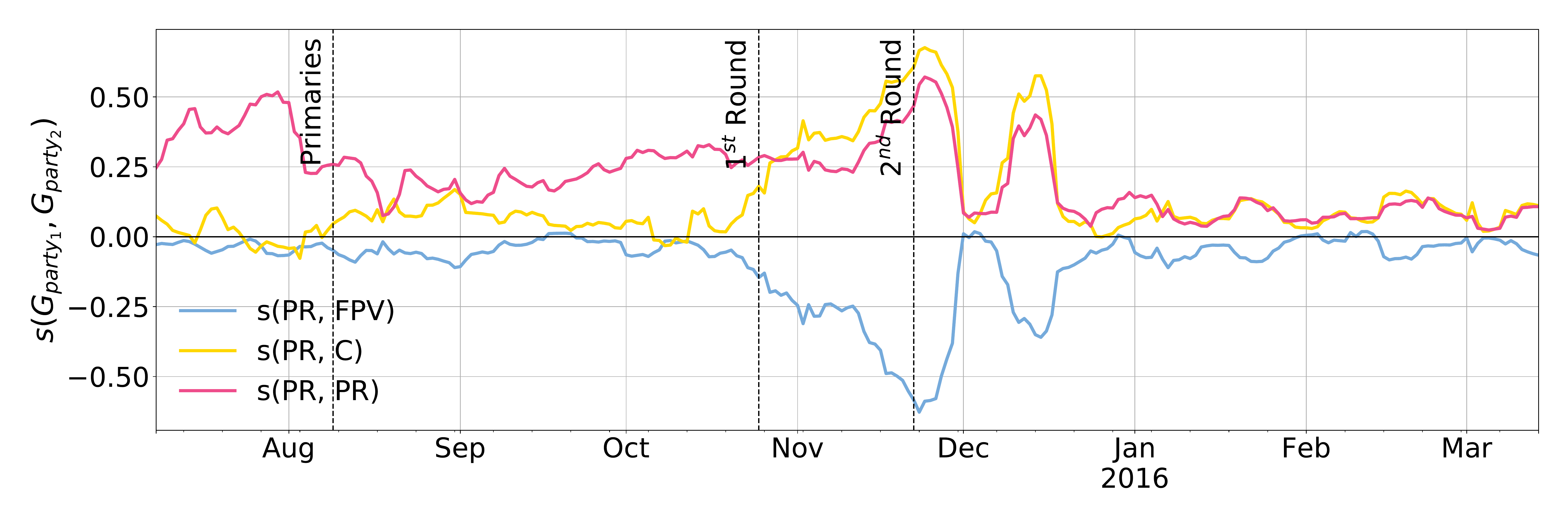}
  \includegraphics[width=16cm]{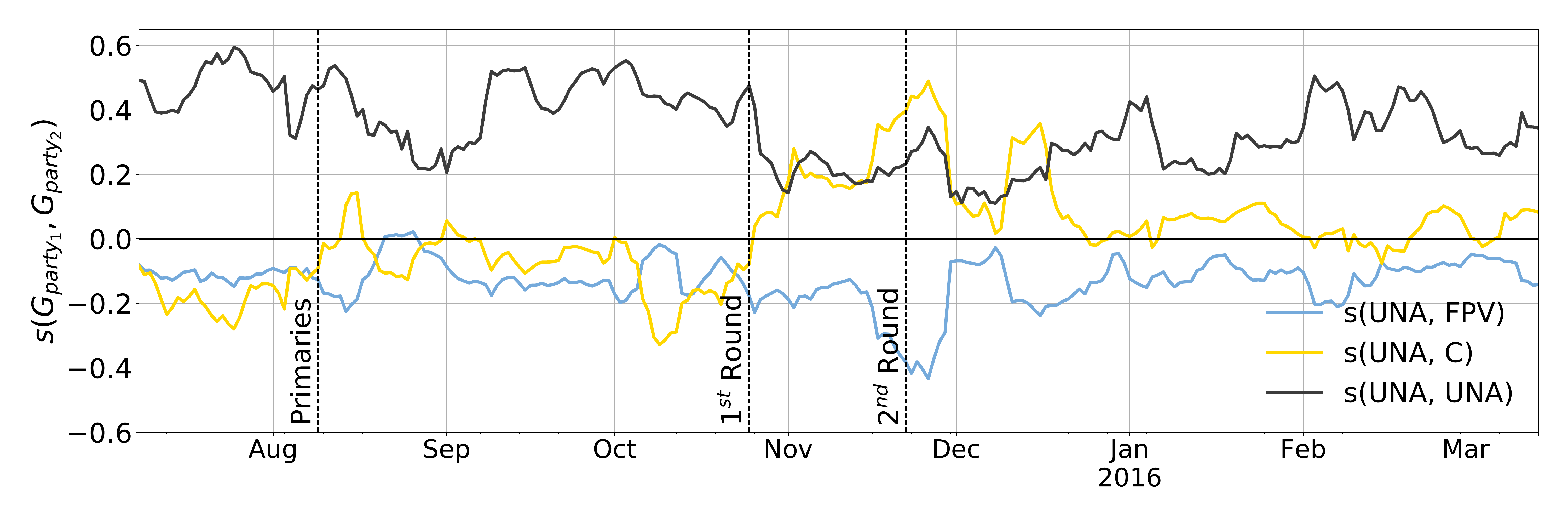}

\caption{\textbf{ Cross-similarities between  each of the two smaller parties and the two major ones in 2015: C (in yellow) and FPV (in light blue)}. In each panel the self-similarity of each minority party is plotted for comparison.
Top: Cross-similarity of PR against the two main parties.; 
Bottom: Cross-similarity of \textit{UNA} against the two main parties.}
\label{2015_cross-sim}
\end{figure*}

The most interesting feature of this electoral period is revealed by the  cross-similarities between each one of the two smaller parties against the two leaders. Figure~\ref{2015_cross-sim} shows the cross-similarity of the two small parties, PR (top) and UNA (bottom) with the two leaders, FPV (blue curve) and C (yellow curve), along with the self-similarity of the corresponding  small party, for comparison. This analysis shows a small positive (negative) cross-similarity of the PR supporters  with the C (FPV) party, compared to their own self-similarity. This  reveals than out of the electoral period, the interests of the PR supporters have little overlap with those  of the dominant parties. However as the first round approaches the behaviour of the PR changes and they clearly show a community of interests with the supporters of C.

A similar behaviour, though more enhanced, happens with the UNA supporters. With a negative cross-similarity against both major parties before the elections, it is between the two rounds that  the  cross-similarity with the C party suddenly strongly increases, showing a  clear choice made by  the UNA supporters. This change in the alignment of the UNA between the two rounds turned out to be decisive to the 2015 election results and the victory of C party.

\section{Summary and Conclusions}


We have performed a  study of the dynamics of opinion during an electoral process, based on data obtained from the micro-blogging platform Twitter. While this subject has been explored in several works ~\cite{tumasjan2010predicting, ma2013predicting,ahmed20162014,caldarelli2014multi,varol2014evolution,zhang2015event, bovet2018validation, gaumont2018reconstruction}, here we apply a different, user-centered, perspective  of the discussions that are taking place in the platform. Most previous works on the subject define a set of keywords, hashtags or mentioned users (e.g., political candidates) to be tracked, and thus they obtain a dataset of tweets which are inherently political. Instead, by defining a set of seed users and capturing all the content that their followers generate, we have information about the evolution of the users' opinion on different topics, and we  are not only restricted to a subset of their tweets.


Following Ref.~\cite{cardodo2019topical}, the topics to study are not set a priori, but emerge from the community structure of a semantic network. This network is built with the assumption that two hashtags used in the same tweet carry some semantic relationship. The disclosed communities provide a representation of the opinion of each user in a multidimensional topic space.
In this work we add the temporal dimension to the topic vectors, and therefore we are able to study the dynamics of the opinion of party supporters during the electoral campaign with great detail.

As discussed in the introduction, the known biases of the population using Twitter  to foster political discussions, which lead among others, to an over representation of an urban male population~\cite{vaccari2013social, barbera2015understanding} hampers the possibility of prediction of electoral results.
Instead, we show here how we can follow  the evolution of  political opinion through the  different stages of an electoral period.  The case of the 2015 elections in Argentina shows that our method captures the details of the reshaping dynamics of the opinion that was decisive to overturn the results within the two rounds of the election.

Although we cannot expect to predict the outcome of an election, one could still attempt to detect massive opinion changes on \textit{real time}. In this respect,
it is worthwhile  recalling some technical details in order to understand the possibilities and the  limitations of the method developed here.
  In this work, the topics are determined by the community analysis of an {\it aggregated} semantic network, meaning that it has been built using the tweets collected during the whole electoral period.
  Tracking the semantic network in real time has the drawback of starting with a small network, with  new hashtags entering as time evolves, which could hamper the correct initial determination of the topics by lack of data.
 A compromise situation would be to start by a semantic network  aggregated during some initial period, in order to set the terms of the public discussion. From that starting point, one could  then incorporate the new hashtags to the existing semantic network, following  a sliding time window. In this way, the topics could  be recalculated and the analysis of the similarities could be performed {\it almost} in real time with just a small lag.  It is expected that the more hashtags enter the semantic network the more accurate will be  the opinion landscape mapping. This assumption lays on
  the implicit hypothesis
 that the system tends to a steady (or meta-stable) state compatible with the aggregated network. However, this may not always be the case. It is easy to imagine that some event, like the present Covid19 pandemic, introduces at some point in time an important, short-time scale, modification of the semantic network. In such cases, integrating new hashtags as described above, may  capture extreme patterns that could be caused by the appearance of a rare event triggering the modification of the structure of the semantic network, almost in real time. This work is in progress.

\section*{Acknowledgements}
This work was done in the framework of the  T-AP Digging into Data OpLaDyn Project, J.I.A.H. and M.G.B. acknowledge the financial support of UBACyT-2018 20020170100421BA, L.H. and D.K.  acknowledge the financial support of 2016–147 ANR OPLADYN TAP-DD2016 and MGB acknowledges the support from the IAS (CY-Cergy-Paris University). The authors acknowledge the former students Facundo Guerrero, Rodrigo de Rosa and Marcos Schapira (Facultad de Ingeniería, Universidad de Buenos Aires) for their contribution to the development of the online platform~\cite{platform}.

\section*{Author contributions}
JIAH, MGB and LH conceived the research project. TMR and MGB wrote the code for extracting the hashtag network, computing the matrices and the evolving similarities, and produced the figures. MGB and LH wrote the manuscript. JIAH, MGB, LH and DK analyzed and discussed the results. All authors contributed in proofreading and discussing the manuscript.


\section*{Competing interests}
The authors declare no competing interests.

\bibliographystyle{apalike}

\newpage

\section*{Supplementary Information}

\subsection{Data description and processing}

\subsubsection{Detail on the collected tweets}
\label{SI_tweets}

\begin{figure*}[h!]
 \centering
 \includegraphics[width=17cm]{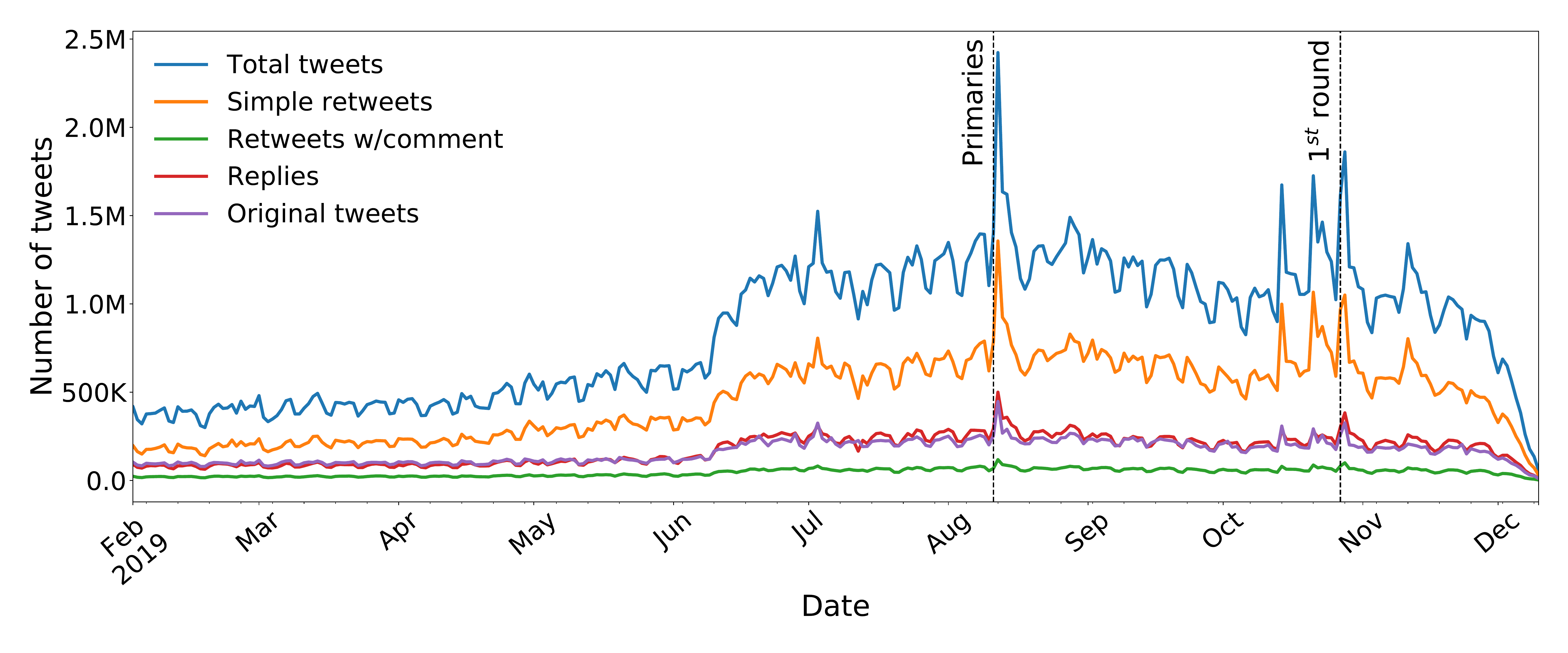}
\caption{Daily number of tweets captured from active Argentinian users (AAU) during the 2019 election. Black dashed lines indicate the dates of the primaries and the elections.}
\label{fig_number_tweets}
\end{figure*}

Figure~\ref{fig_number_tweets} shows  the evolution of the number of tweets collected per day during the  2019 election period, along with their discrimination in terms of new tweets, retweets (simple and with comment) and replies to other tweets.

The daily number of tweets in the capture reveals the standard weekly patterns in the usage of the platform, with less activity during the weekends. The high peaks just before the primaries of August 9th and during the three mondays close to the main election reveal the interest of our users in the political events.

We classified the tweets into $4$ groups, using the following criterion:
\begin{itemize}
\item \textbf{Replies:} tweets having the \texttt{is\_reply} flag active according to the Twitter API.
\item \textbf{Simple retweets:} tweets having the \texttt{is\_retweet} flag active according to the Twitter API.
\item \textbf{Retweets with comment (quote tweets):} tweets having the \texttt{is\_quote} flag active according to the Twitter API, but have the \texttt{is\_reply} and \texttt{is\_retweet} flags down.
\item \textbf{Original tweets:} tweets having both the \texttt{is\_quote}, \texttt{is\_reply} and \texttt{is\_retweet} flags down.
\end{itemize}

We observe that almost $55\%$ of the tweets captured are simple retweets, while almost $20\%$ are original tweets.

\begin{figure*}
  \centering
  \includegraphics[width=8cm]{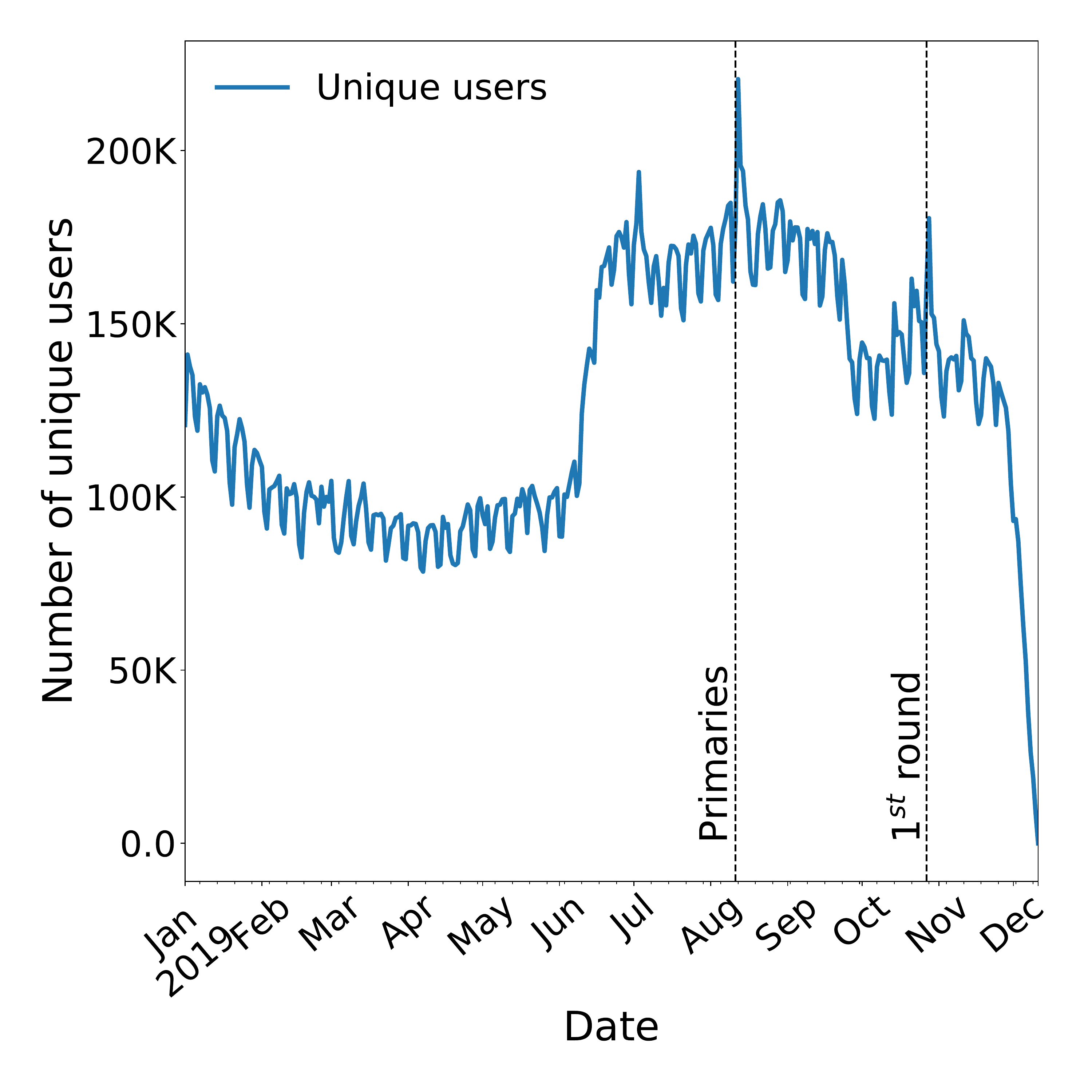}
  \includegraphics[width=8cm]{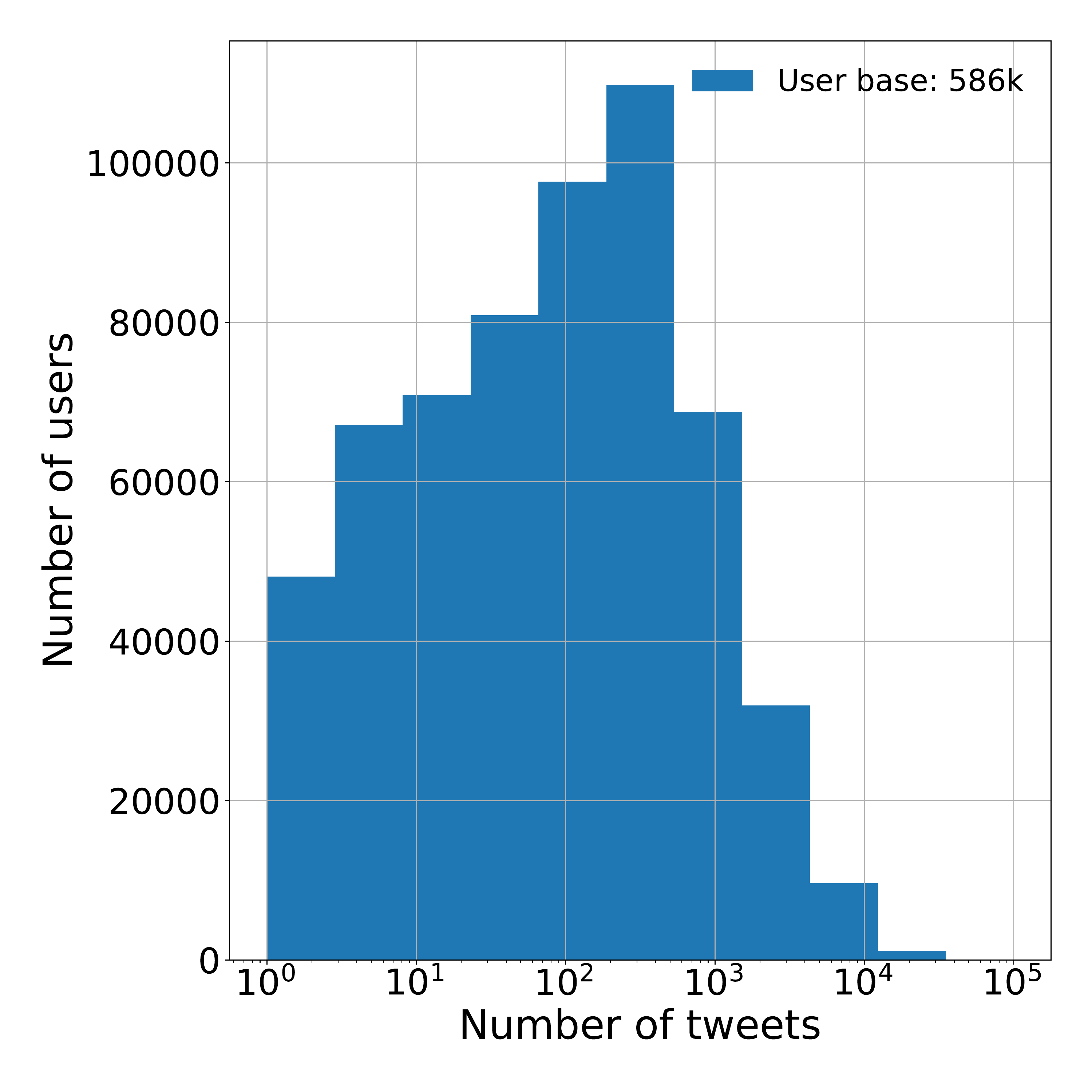}
\caption{
  \textit{(Left)} Number of argentinian users from our base active in the platform during each day.
  \textit{(Right)} Histogram of the total number of tweets per user in the 2019 election.
}
  \label{fig_SI_1}
\end{figure*}

Each day, an average of $126k$ argentinian users in our dataset showed some activity in the platform (see Figure~\ref{fig_SI_1}, left). This represents the $21\%$ of our the base of active argentinian users (AAU). The average number of tweets captured per user was $477$ or, in other words, $1.4$ average tweets per user per day (Figure~\ref{fig_SI_1}, right).

\subsubsection{User affiliation}
\label{SI_user_affiliation}

The affiliation of users to some political party was inferred from their follower relationships and retweets from the candidates' accounts. We consider two scenarios:
\begin{itemize}
    \item The user did not retweet from any candidate: In this cae, if he only follows candidates from a single party, then we infer that the user supports that party.
    \item The user retweeted from at least one candidate: In this case, we count his retweets from each party, and we normalize them so that they add up to $1$. If some party gets a proportion of at least $0.75$, then we consider that the user supports that party.

\end{itemize}
Users that do not fulfill any of these conditions are not considered in any of our similarity analysis. The political affiliation of the users was determined  during the pre-election period, before the primaries.

\subsubsection{Geographical and gender representativity}

\begin{figure*}[t]
  \centering
  \includegraphics[width=16cm]{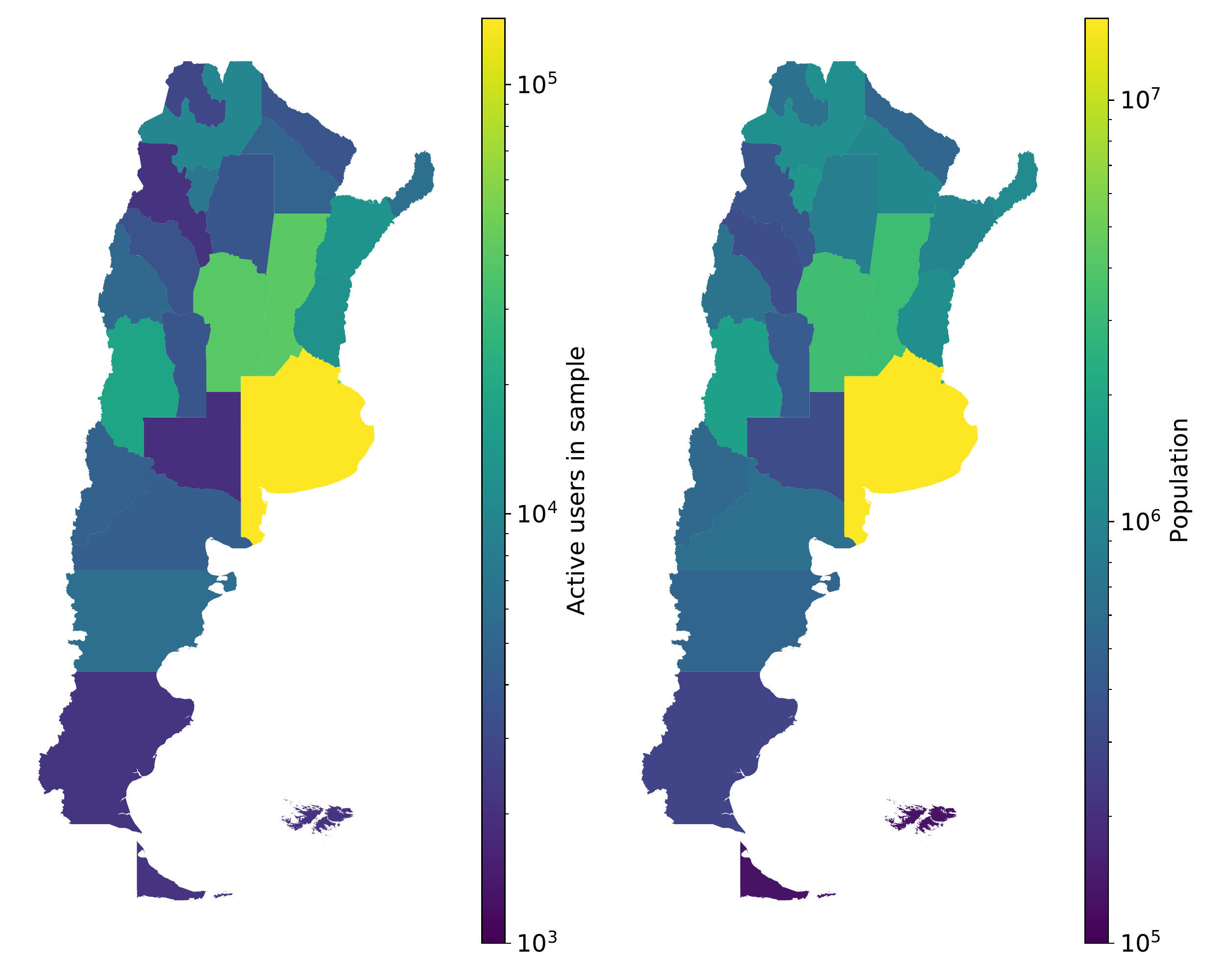}
\caption{
  \textbf{Users representativity at the province level.} Number of users per province in our sample in the 2019 election \textit{(left)}, as compared to the general population distribution in Argentina \textit{(right)}. The official population data was obtained from the IGN (\textit{Instituto Geográfico Nacional}).
}
  \label{fig_map_provinces}
\end{figure*}

Each user's location was computed by combining the location informed in the user profile with data from GeoNames (\url{https://www.geonames.org/}). In the 2019 capture, for 392k users we could determine a specific location inside the country (either city or province), and for those we plotted their geographical distribution. Figure~\ref{fig_map_provinces} shows that the capture has a fair geographical representativity at the province level. A similar representativity was found for the 2015 capture.

Twitter does not provide gender information in the user profiles. Thus, we analyzed the names in the profile by comparing them to a list of historical names from Argentina (\url{https://data.buenosaires.gob.ar/dataset/nombres)}. In the 2019 capture, $32\%$ of the profile names did not match any registered name (e.g., some users do not write their names but use aliases and/or emojis). For the remaining users (383k users), we found a $43\%$ of females and $57\%$ of males. In the 2015 capture we found a $55\%$ of males and a $45\%$ of females, after filtering a $32\%$ of users with unknown gender.

\subsubsection{Hashtag usage}
\label{SI_hashtags}

\begin{figure*}
  \centering
  \includegraphics[width=8cm]{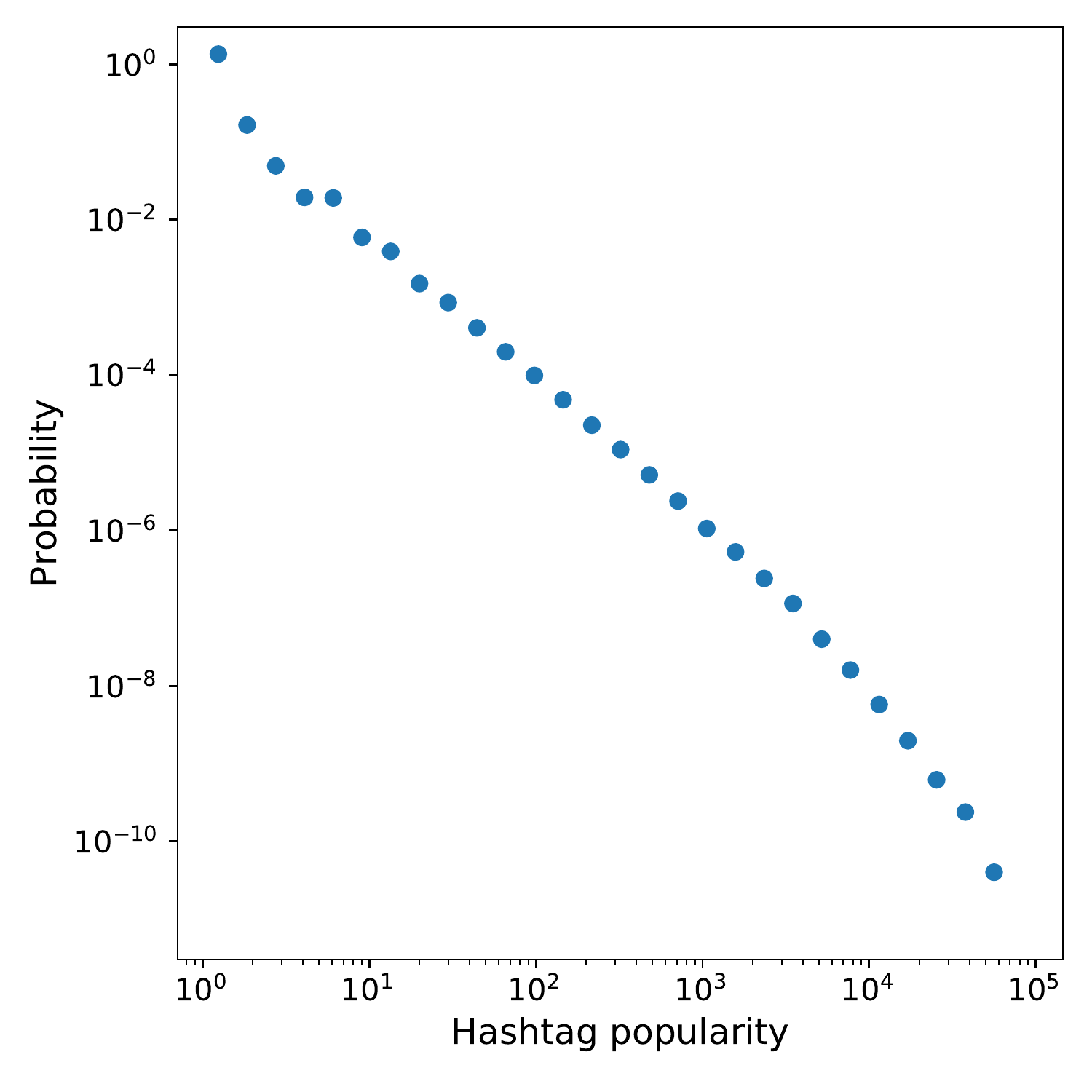}
  \includegraphics[width=8cm]{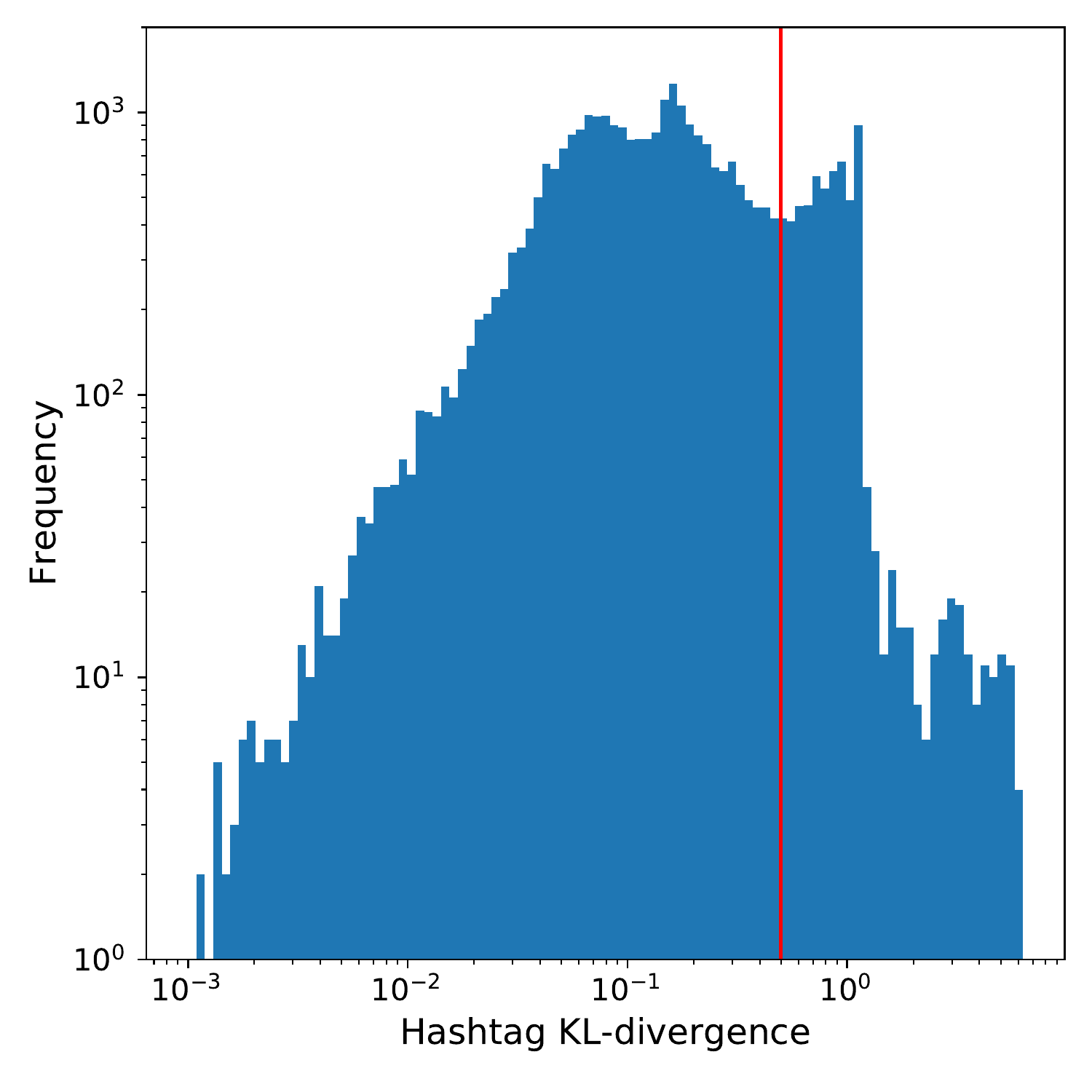}
\caption{
  \textit{(Left)} Histogram of the number of unique users adopting each recorded hashtag during the 2019 election.
  \textit{(Right)} Histogram of the KL divergence between hashtags' and users' distributions across parties during the 2019 election. The red vertical line represents the threshold for hashtag classification as political ($\geq 0.5$) or apolitical.
}
\label{fig_hashtag_popularity}
\end{figure*}

Hashtags are used in Twitter as a tagging convention to associate \textit{tweets} with events, concepts or contexts. We found that $~14\%$ of the \textit{tweets} contain at least one hashtag. Some of these hashtags become very popular while others go unnoticed; the probability distribution of hashtag usage follows a heavy tail (Figure~\ref{fig_hashtag_popularity}, \textit{left}), with $66\%$ of the hashtags used by only one person, and $2,600$ hashtags adopted by at least $1,000$ users.

\textit{Detecting political hashtags.}
Some hashtags have a stronger political meaning than others.
We designed a method to automatically identify political hashtags based on their usage by supporters of the different parties. In effect, we assessed the correlation between hashtag usage and political affiliation in terms of the Kullback-Leibler divergence (\textit{relative entropy}): given the empirical usage of each hashtag $h$ by users supporting each political party, we define a generalized Bernoulli distribution $P_h$ for each hashtag (represented as a vector containing the proportion of usages of hashtag $h$ in each party with respect to all users), and a generalized Bernoulli distribution $Q$ measuring the distribution of users across parties (i.e., the proportion of users supporting each party). Then, we compute the relative entropy $D_{KL}(P_h||Q)$ between these distributions for each hashtag $h$:
\begin{equation}
D_{KL}(h) = \sum_{i=1}^{N_C} P_h(i) \cdot \log_{2}{\left[\frac{P_h(i)}{Q(i)}\right]},
\end{equation}

where $N_C$ represents the number of parties. The idea behind this method is that apolitical hashtags $h$ will be used independently of party affiliation, and their distribution will be similar to that of the party sizes, implying that $D_{KL}(h)\approx 0$, while strongly political hashtags will have higher relative entropy values.
The observed distribution of these relative entropies for the 2019 election is shown in Figure~\ref{fig_hashtag_popularity} \textit{(right)}, together with the threshold of $0.5$ that we used to classify hashtags into political or apolitical. Hashtags above that threshold will be considered political.

\textit{Weighted hashtag network.}
This network was build by computing the number of unique users that ever used each pair of hashtags together in a tweet. Counting the number of unique users instead of the number of tweets allows for filtering the behavior of users that continually post similar content, and is a way of controlling for bot-generated content. The semantic association of $2$ hashtags appearing in the same tweet is reinforced by the fact that many people use them together. We removed links whose weight is below a threshold of 5.

Table~\ref{hashtag_network_stats} shows a brief description of these graphs.

\begin{table}
\centering
\caption{Weighted hashtag networks description.}
\label{hashtag_network_stats}
\begin{tabular}{lll}
\hline\noalign{\smallskip}
Hashtag network & 2015 election & 2019 election  \\
\noalign{\smallskip}\hline\noalign{\smallskip}
\textit{After threshold} \\
Nodes & $13767$ & $31405$ \\
Edges & $50516$ & $106366$ \\
Communities (OSLOM) & $1863$ & $2139$ \\
\noalign{\smallskip}\hline
\end{tabular}
\end{table}

\subsubsection{Robustness to fake accounts and bots}

We made an effort to mitigate the impact of bots and fake accounts during the design of the method, by following these strategies:

\begin{itemize}
    \item As mentioned before, in the semantic network built by hashtag co-occurrence, each user is counted only once. That is, when we count the number of simultaneous usages of a pair of hashtags, we do not count the number of tweets but the number of unique users.
\item When we measure the general usage vectors $T_i$ each user gets the same weight, because his description vector has been normalized. In this way, users with a very high activity do not distort the average behavior.
\end{itemize}

Contrary to other capture methods which are based on keywords, and which are more prone to capturing bot content, our method is based on chosen users (i.e., those who follow some candidate). This reduces the effect that fake accounts could have in our Twitter capture.

\subsection{Graph structure and evolution of selected topics}

Here we show the evolution in the discussion of certain topics during both elections. These topics were selected based on two criteria: \textit{(i)} to show the opinion of Twitter users on some of the hot topics of discussion during each period (as political tensions in neighboring countries, abortion legislation and the defense of human rights) and; \textit{(ii)} to validate the correct affiliation of users into parties by looking at the composition of the main topics that reflect party support.

\begin{figure*}[h!]
  \centering
  \includegraphics[width=8cm]{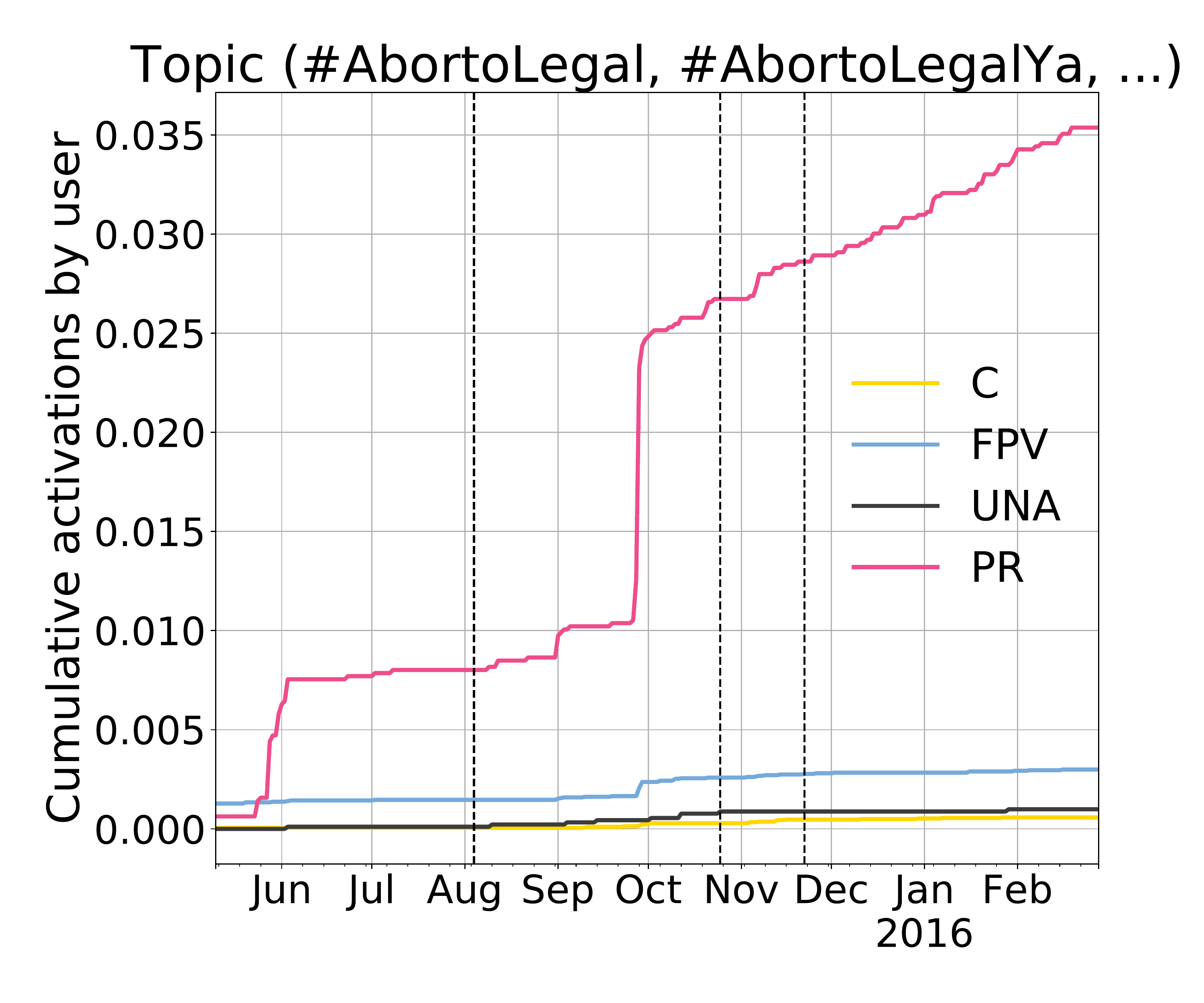}
  \includegraphics[width=8cm]{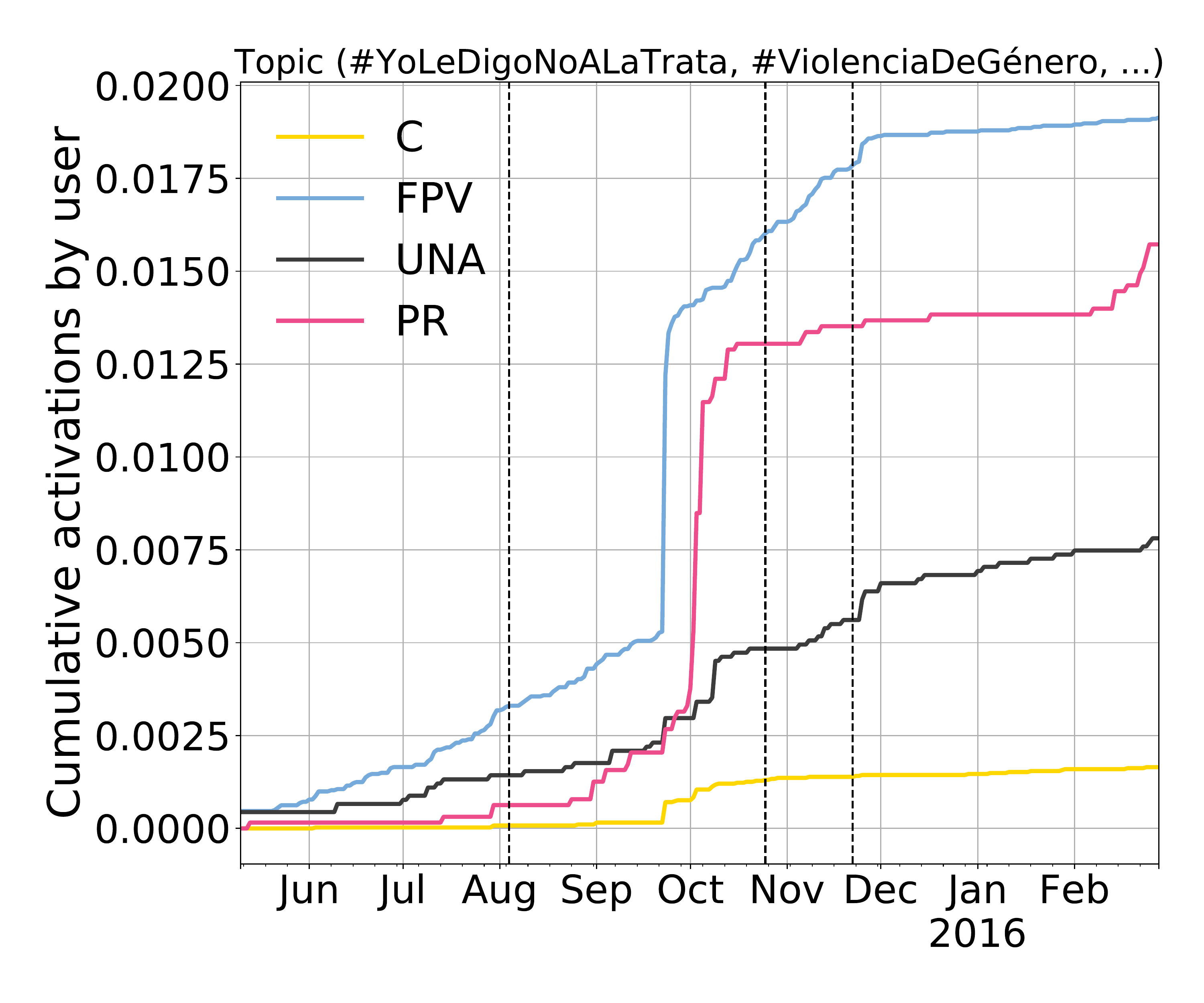}
\caption{\textbf{Hot discussions during the 2015 election. Each panel shows the cumulative usage of the topic by supporters of each party.} \textit{(Left)} Topic with demands for legal abortion. \textit{(Right)} Topic with discussions against gender violence.}
\label{topic_evolution2015_abortion_gender}
\end{figure*}

In Figure ~\ref{topic_evolution2015_abortion_gender} we show the evolution of the discussions on legal abortion and gender violence during the 2015 election. We observe that the discussions on legal abortion (left panel) were mainly conducted by the \textit{Progresistas} party (PR), with a minor participation from one of the two major parties, FPV. This topic was particularly active on September 28th, during the International Safe Abortion Day, and continued from then on. On the right panel, the discussions about gender violence were carried out by the center-left party FPV (in charge) and the PR party. The supporters of the center right party \textit{Cambiemos} (C) were the less active in the topic.

\begin{figure*}[h!]
  \centering
  \includegraphics[width=7.3cm]{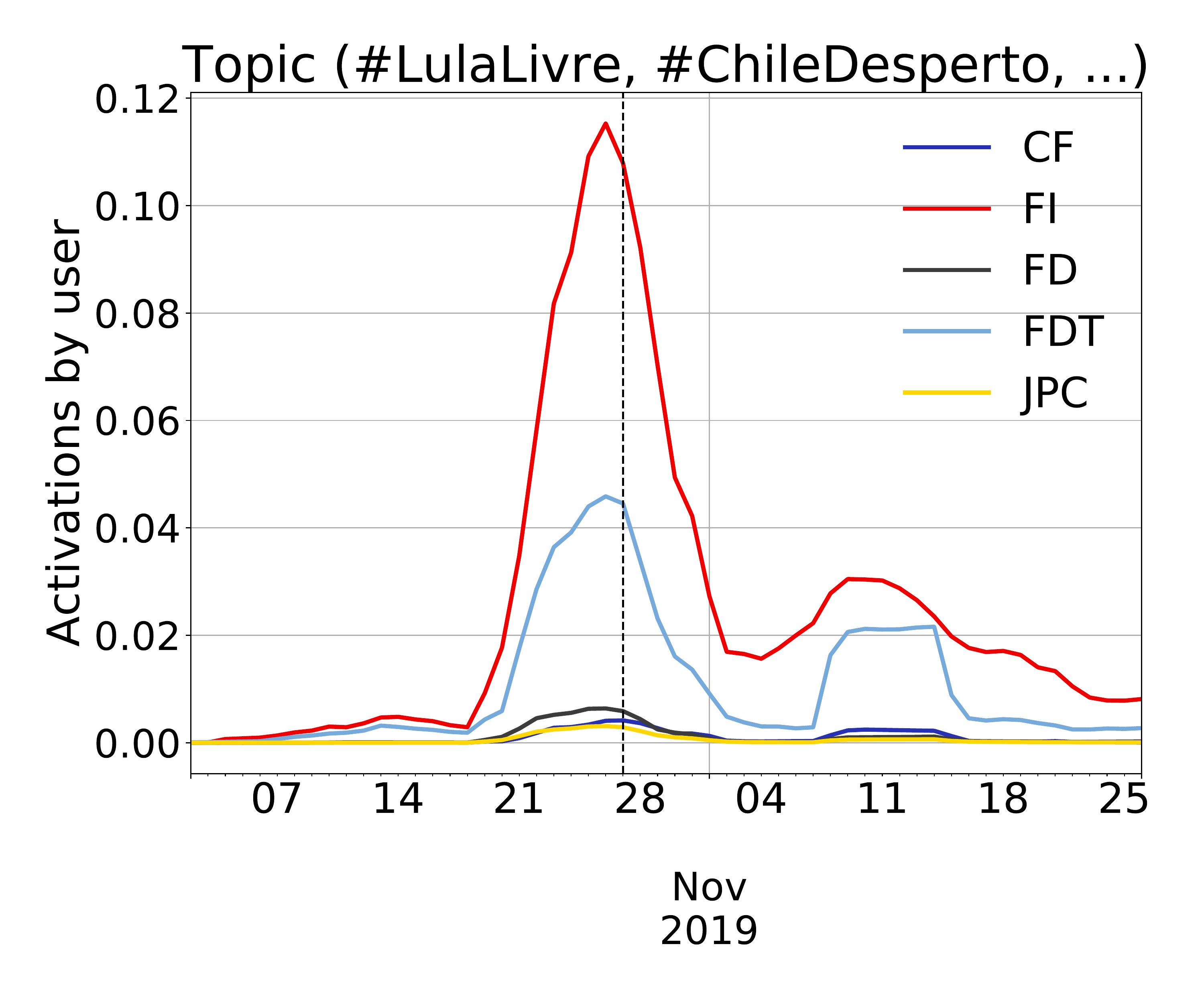}
  \includegraphics[width=7.8cm]{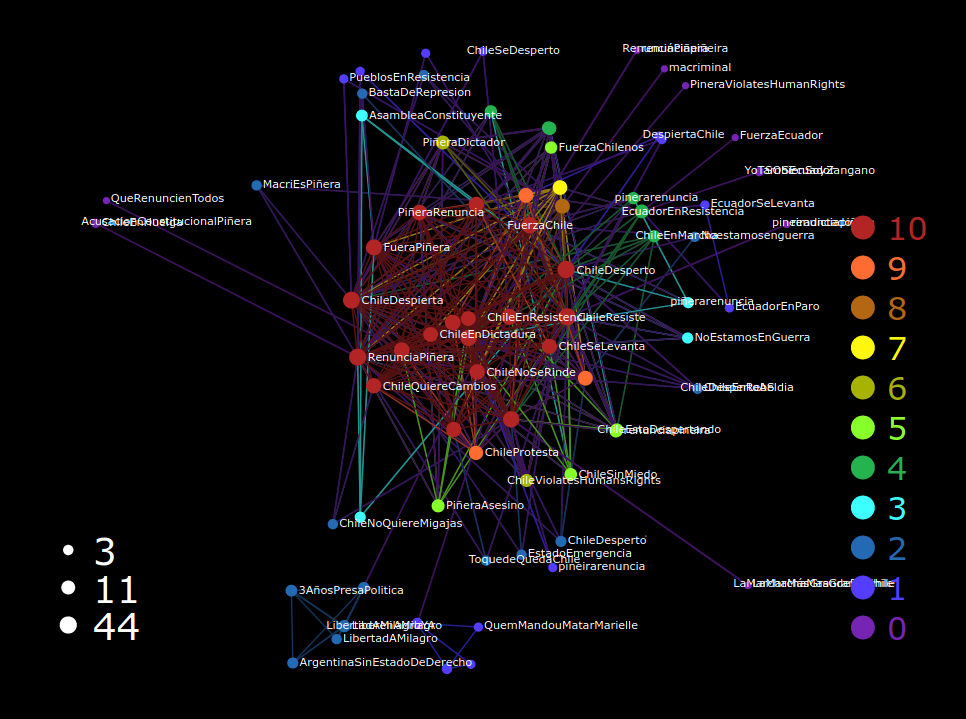}
\caption{
\textbf{Foreign affairs discussions during the 2019 election. Positions regarding social tension in Chile and Brazil.} \textit{(Left)} Time evolution of the topic usage by supporters of the different parties (7-day rolling average). \textit{(Right)} Hashtag composition of the topic. Nodes are arranged according to the $k$-core decomposition of the community graph. The hashtags in this topic (pointing out that `Chile awakened' and `Free Lula') correspond to the interest of the leftist parties, which is reflected by the red and light blue peaks.}
\label{topic_72_evolution2019}
\end{figure*}

\begin{figure*}[h!]
  \centering
  \includegraphics[width=7.2cm]{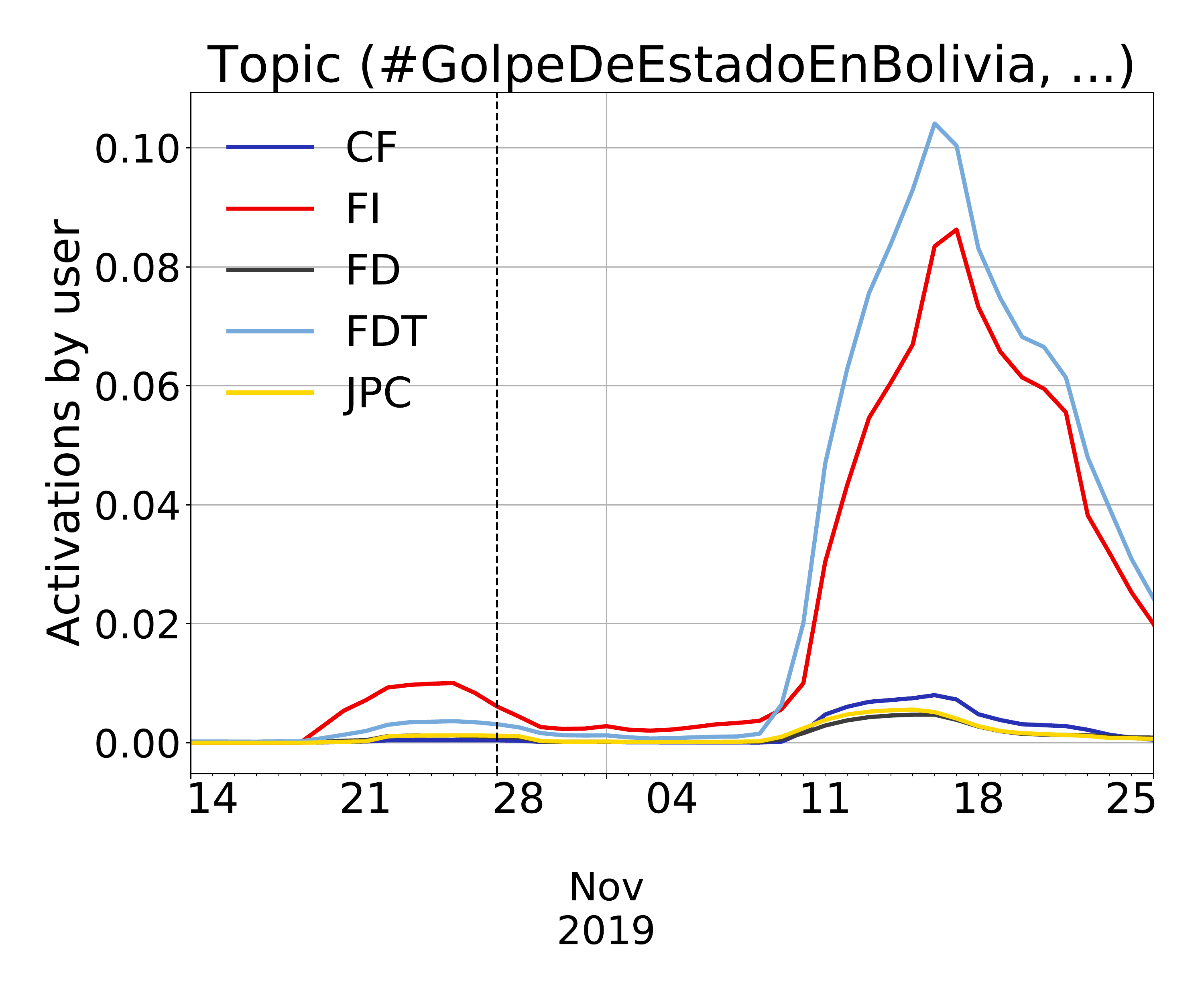}
  \includegraphics[width=8.5cm]{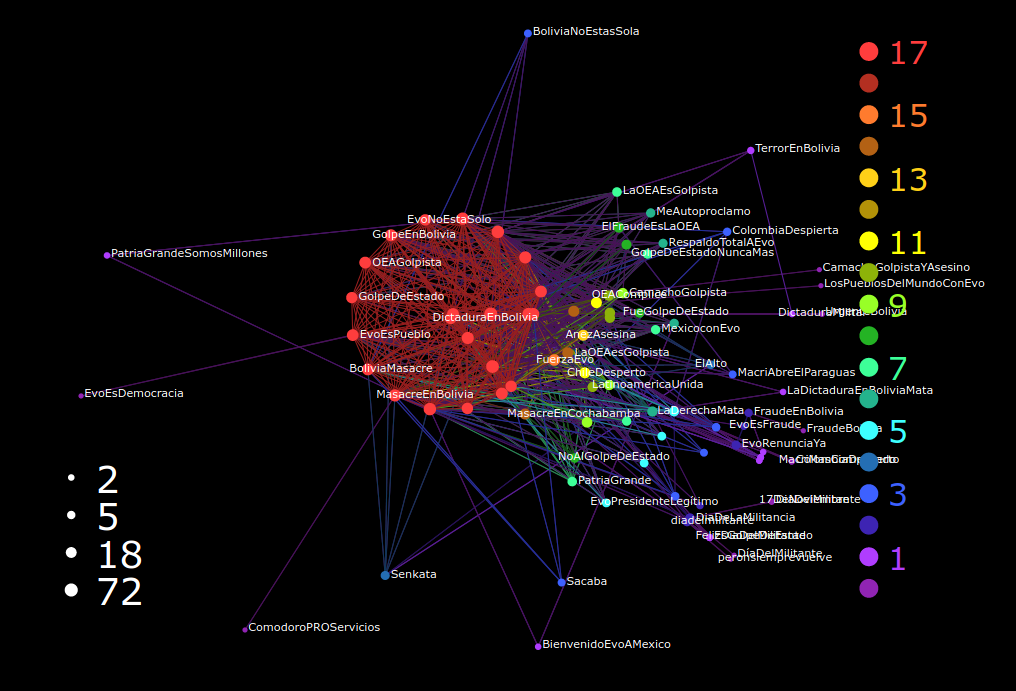}
\caption{\textbf{Foreign affairs discussions during the 2019 election. Positions regarding the \textit{coup d'état} in Bolivia.} \textit{(Left)} Time evolution of the topic usage by supporters of the different parties (7-day rolling average). \textit{(Right)} Hashtag composition of the topic. Nodes are arranged according to the $k$-core decomposition of the community graph. Again, the choice of the hashtags reveals political orientation, as right parties in general did not admit that the events taking place in Bolivia in 2019 corresponded to a \textit{coup d'état}.}
\label{topic_63_evolution2019}
\end{figure*}

Figures~\ref{topic_72_evolution2019} and~\ref{topic_63_evolution2019} show two discussions on foreign affairs during the 2019 election: the social tensions in Chile and Brazil in October 2019, and the \textit{coup d'état} in Bolivia in November 2019. The left-wing parties FI and FDT were the ones most sensitive to these topics, as expected. In the case of Bolivia, it is interesting to point out that, while the small and more leftist FI was the first to raise concerns at the end of October (and as the election first round was taking place), the major left party FDT went ahead during November, with the election already decided. The plots in the right panel of each figure show the main hashtags involved in each case.

\begin{figure*}[h!]
  \centering
  \includegraphics[height=5.7cm]{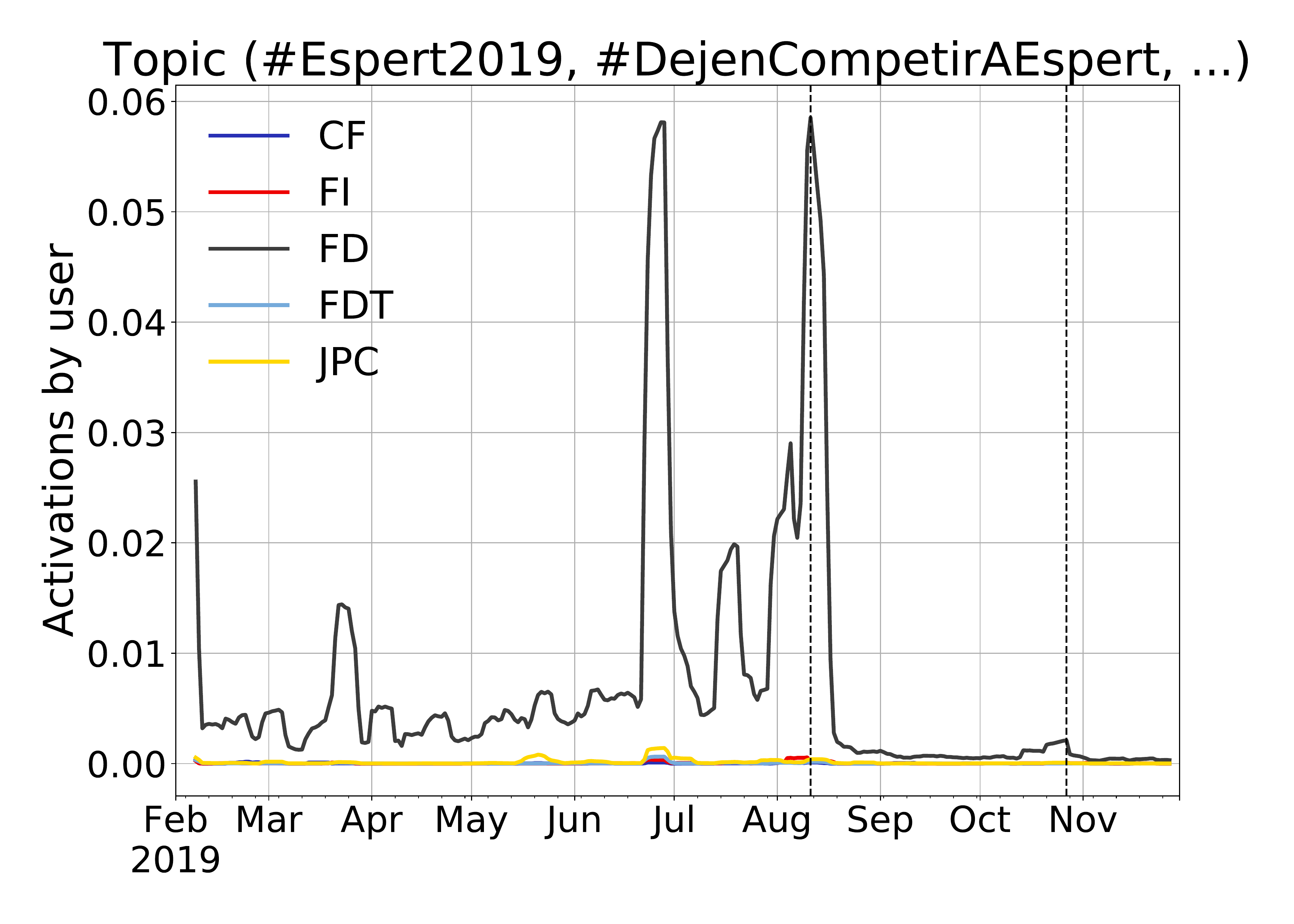}
  \includegraphics[height=6.1cm]{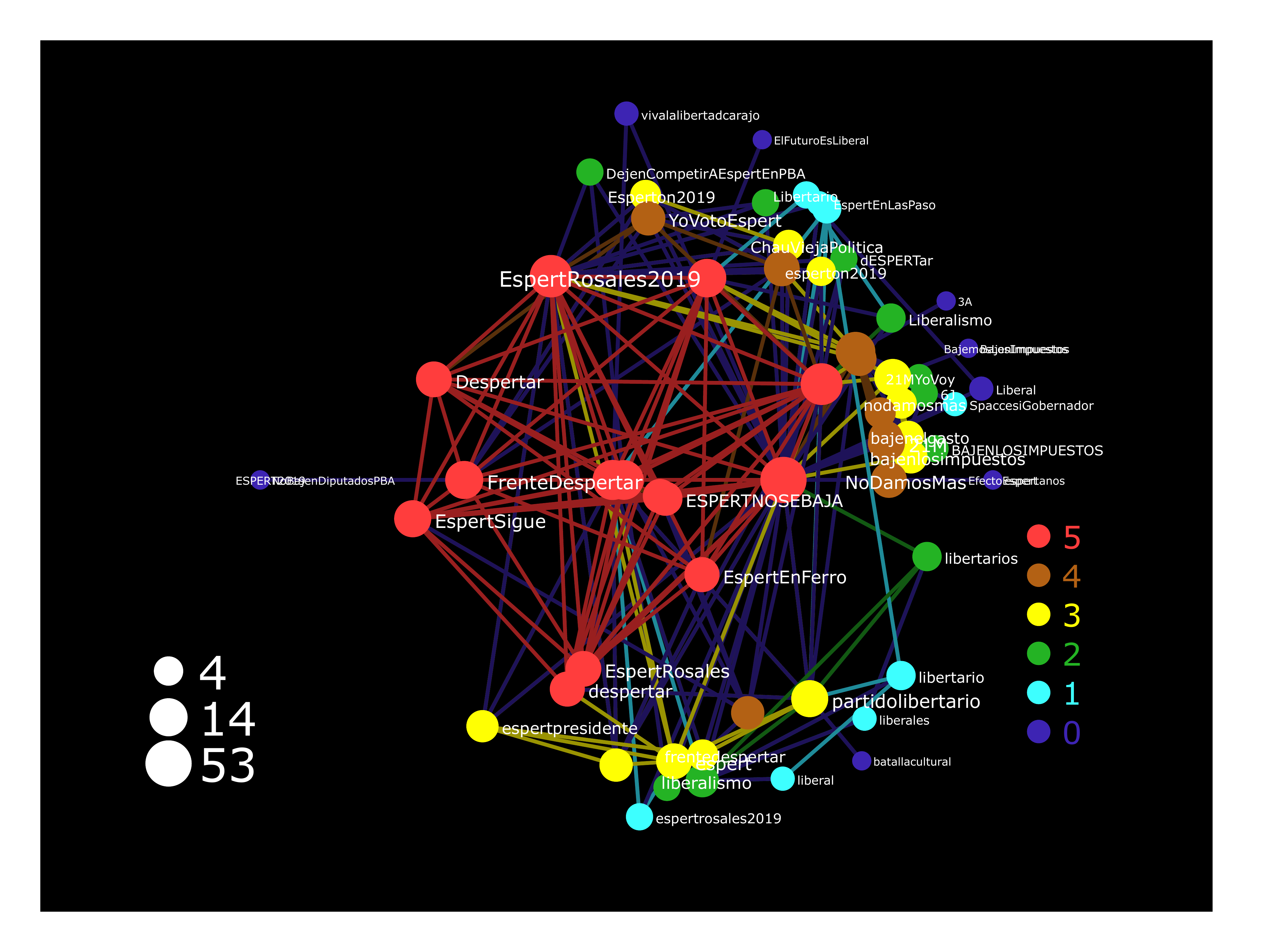}
    \caption{\textbf{Topic expressing the support for the participation of one political actor from the FD party (J.L. Espert) in the 2019 election.} \textit{(Left)} Time evolution of the topic usage by supporters of the different parties (7-day rolling average). \textit{(Right)} Hashtag composition of the topic. Nodes are arranged according to the $k$-core decomposition of the community graph.}
\label{topic_46_evolution2019}
\end{figure*}

\begin{figure*}[h!]
  \centering
  \includegraphics[height=6.4cm]{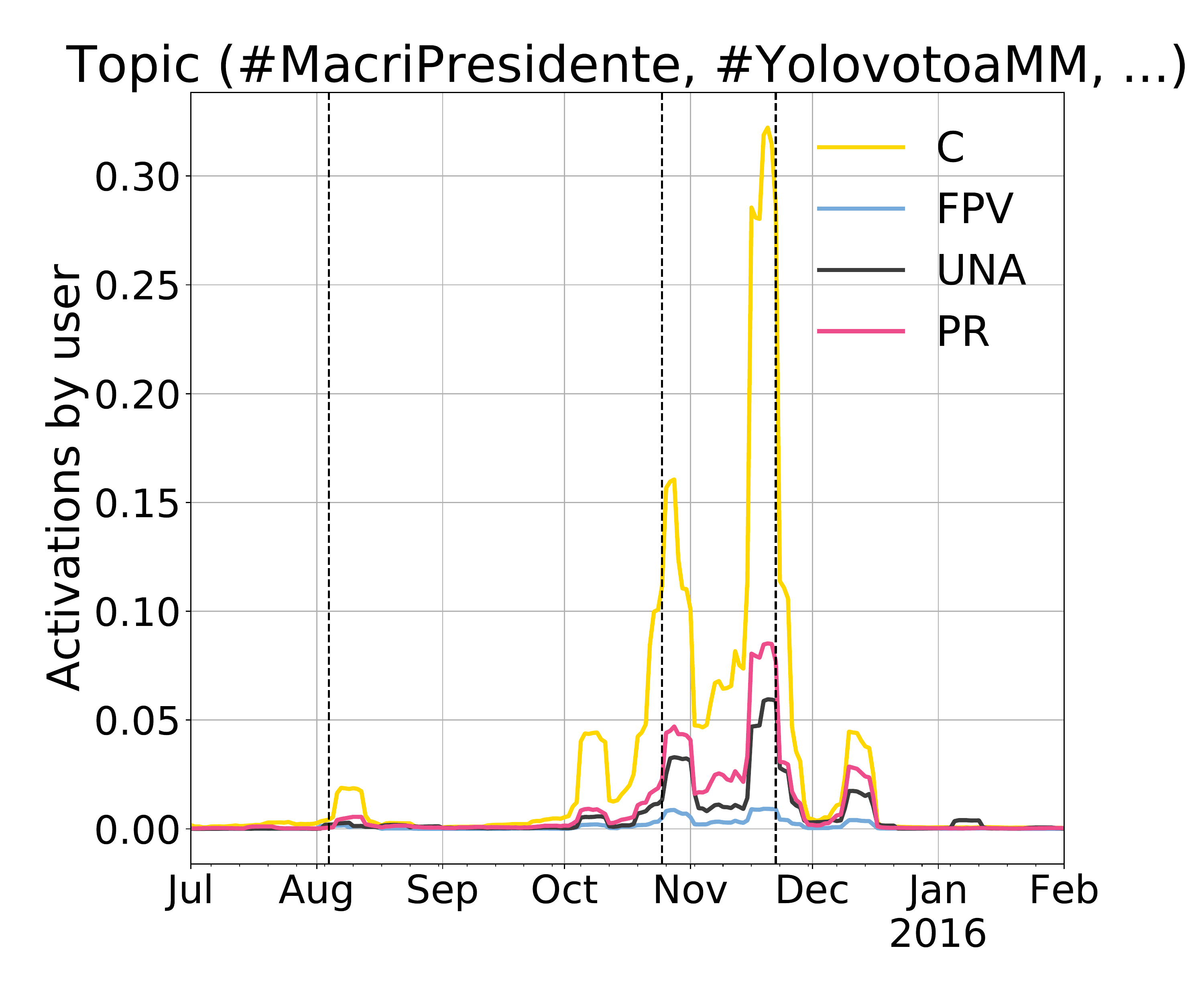}
  \includegraphics[height=6.4cm]{topic_63_2015.pdf}
      \caption{\textbf{Topic expressing the support for the candidate of the C party (M. Macri), who won the 2015 election.} \textit{(Left)} Time evolution of the topic usage by supporters of the different parties (7-day rolling average). \textit{(Right)} Hashtag composition of the topic. Nodes are arranged according to the $k$-core decomposition of the community graph.}
\label{topic_63_evolution2015}
\end{figure*}

\begin{figure*}[h!]
  \centering
  \includegraphics[height=6.4cm]{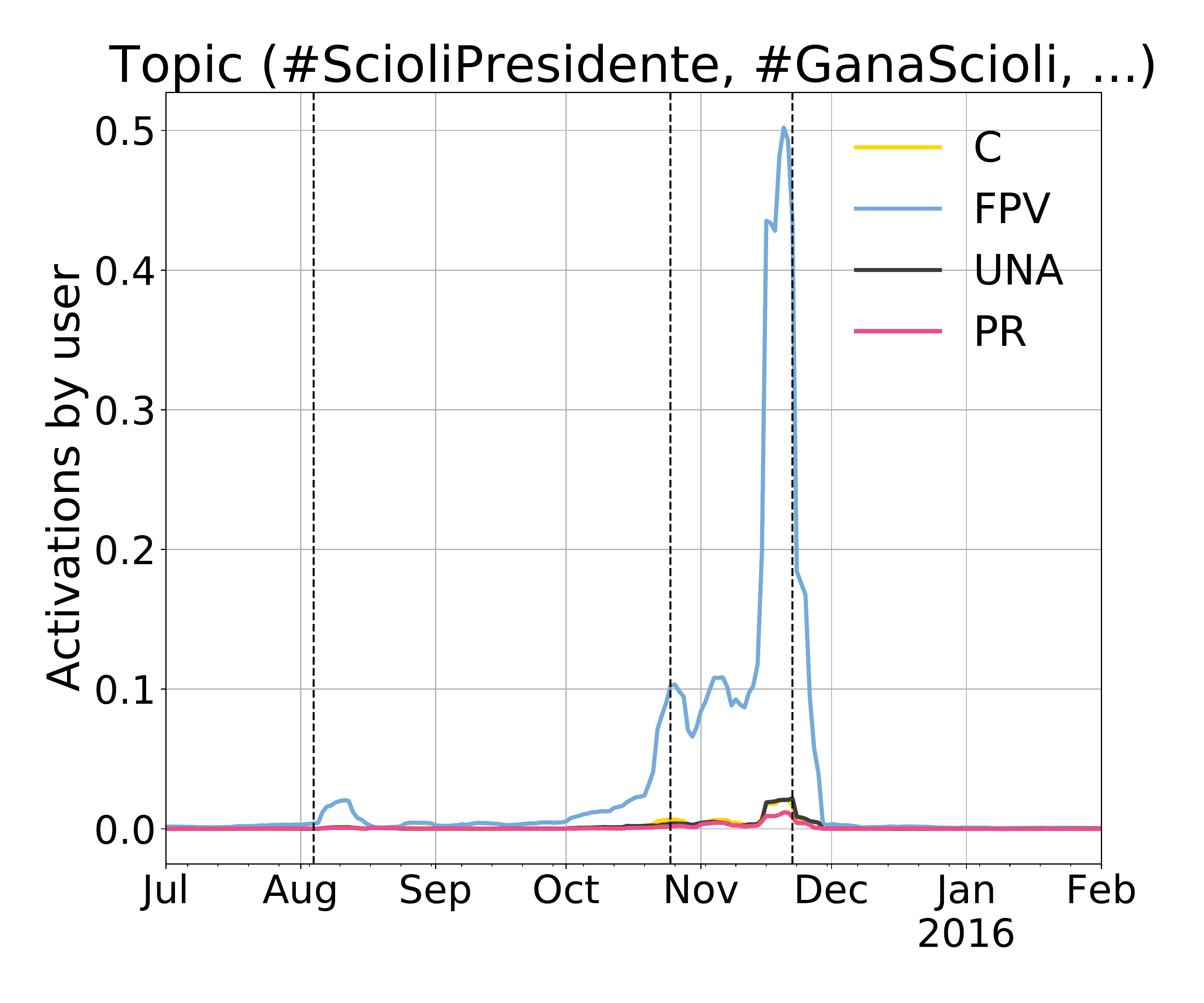}
  \includegraphics[height=6.4cm]{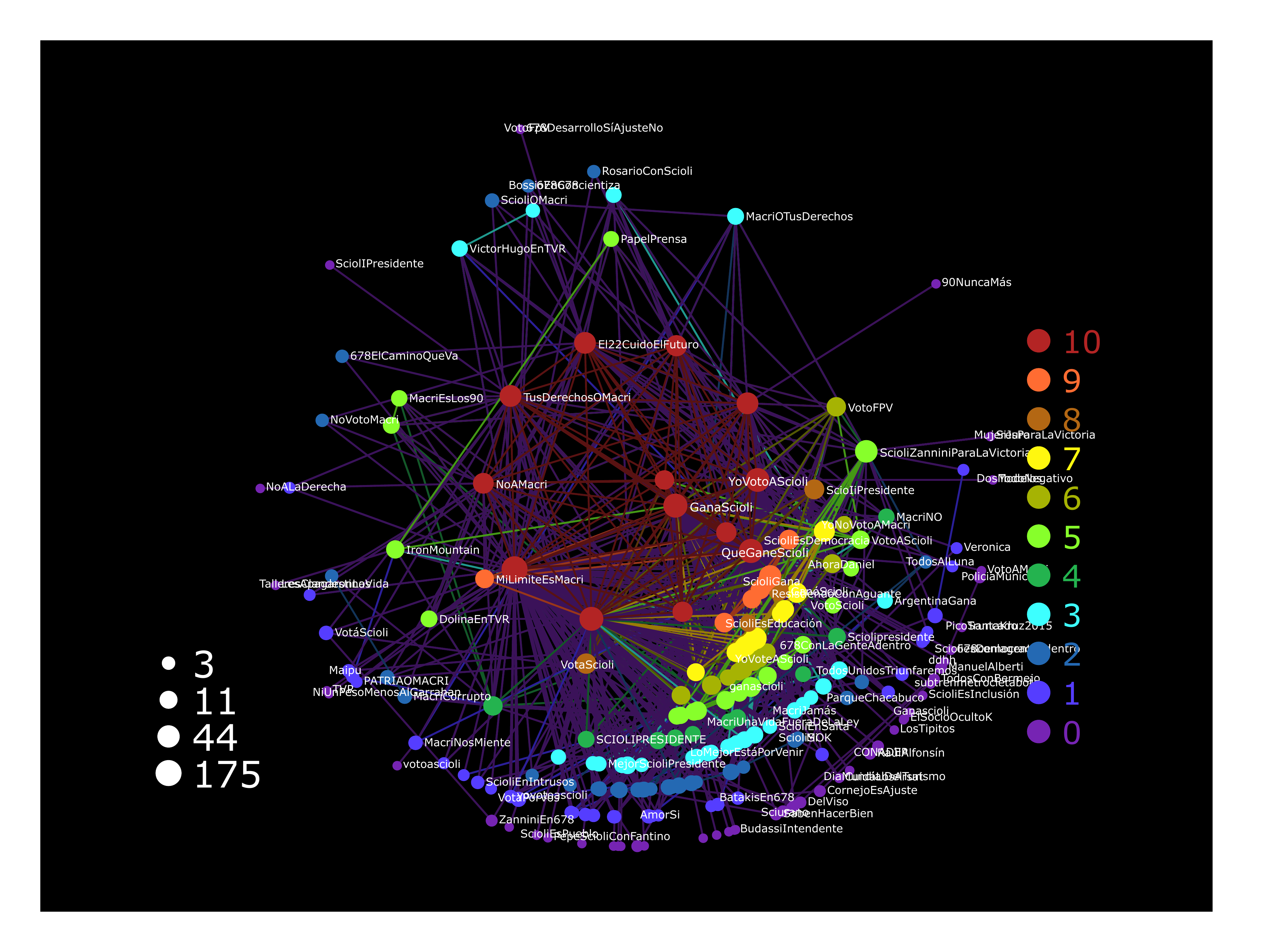}
      \caption{\textbf{Topic expressing the support for the candidate of the FPV (D. Scioli) in the 2015 election.} \textit{(Left)} Time evolution of the topic usage by supporters of the different parties (7-day rolling average). \textit{(Right)} Hashtag composition of the topic. Nodes are arranged according to the $k$-core decomposition.}
\label{topic_80_evolution2015}
\end{figure*}

\begin{figure*}[h!]
  \centering
  \includegraphics[height=5.8cm]{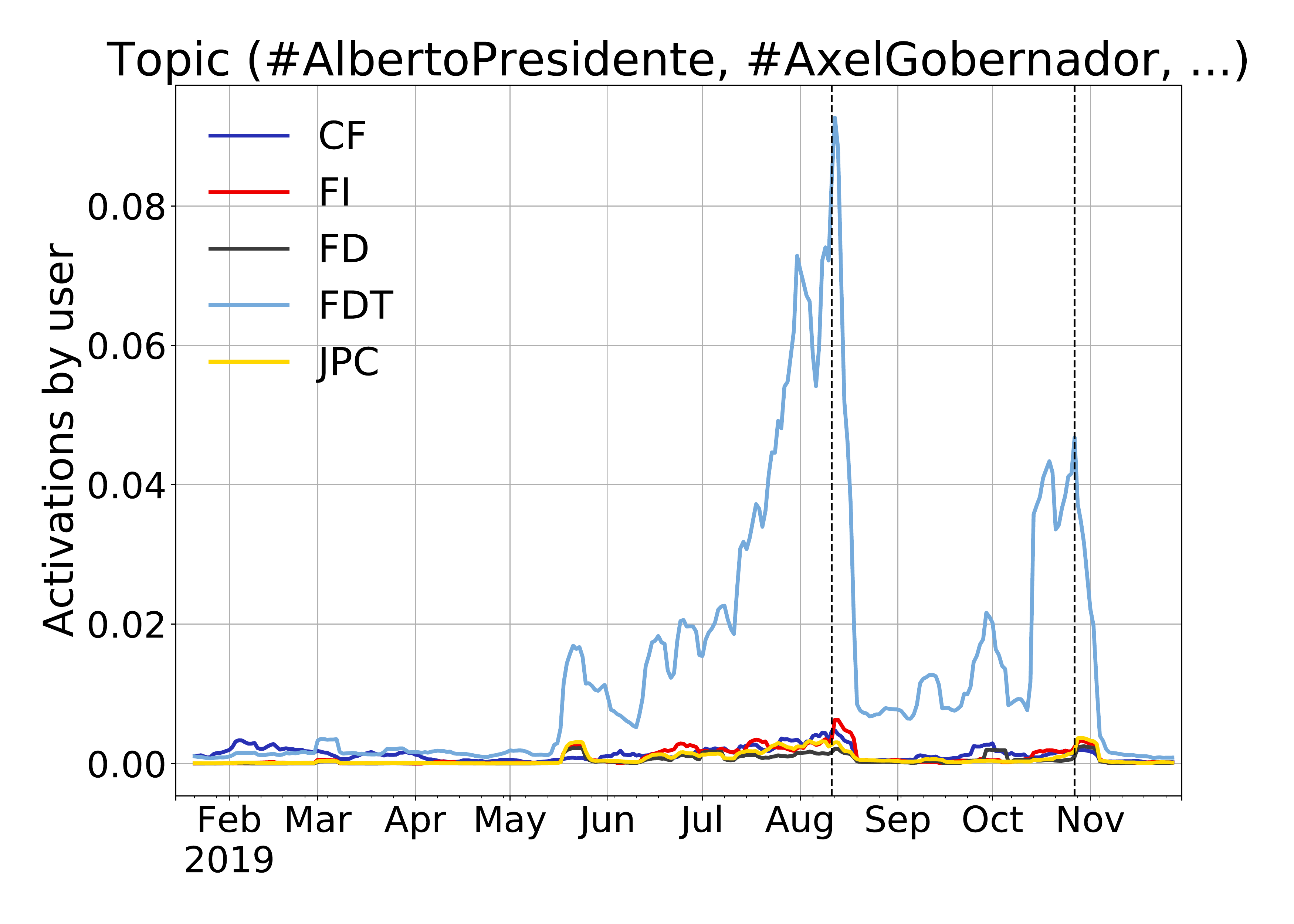}
  \includegraphics[height=6.1cm]{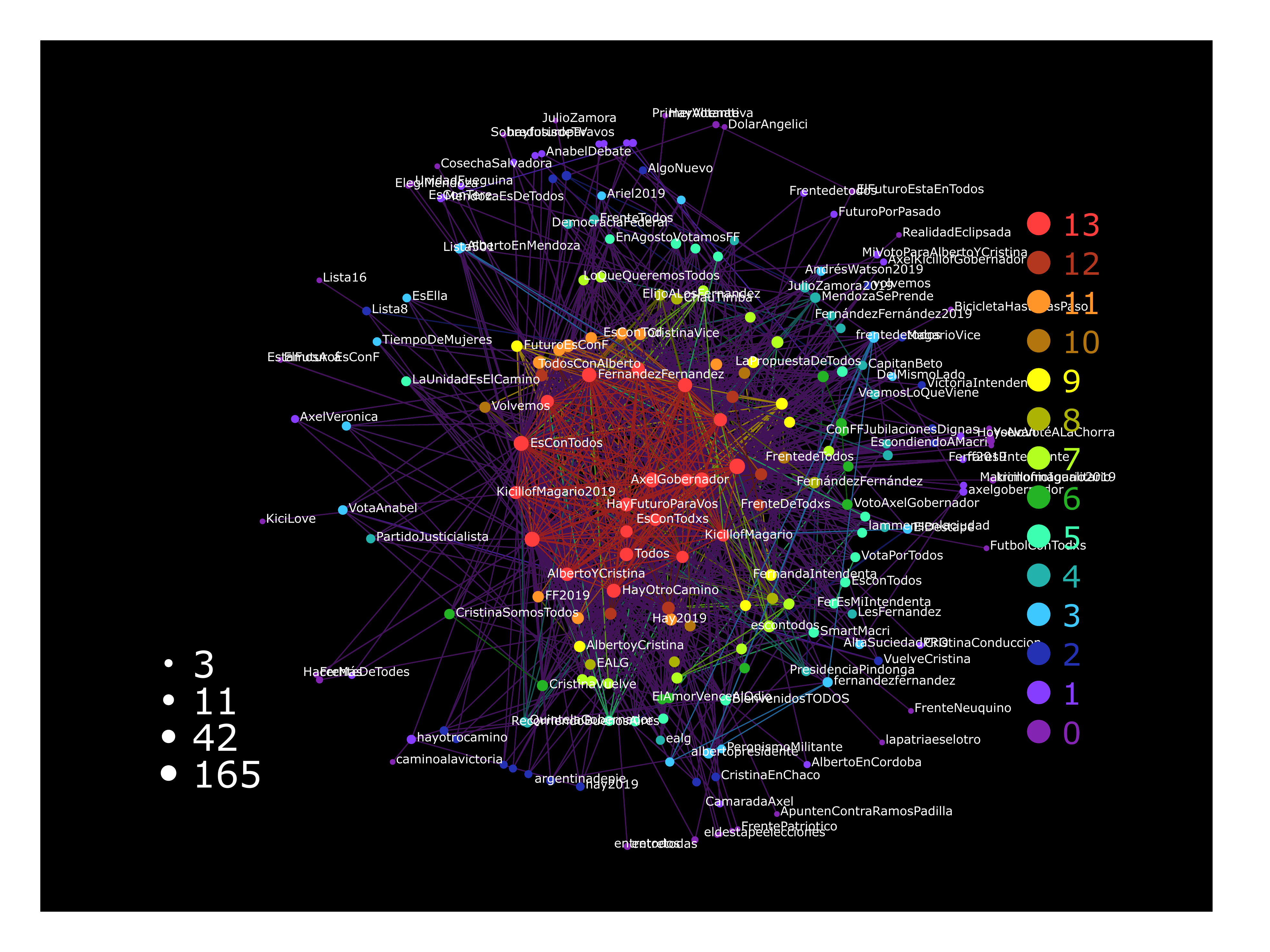}
      \caption{\textbf{Topic expressing the support for the candidate of the FDT party, who won the 2019 election.} \textit{(Left)} Time evolution of the topic usage by supporters of the different parties (7-day rolling average). \textit{(Right)} Hashtag composition of the topic. Nodes are arranged according to the $k$-core decomposition of the community graph.}
\label{topic_64_evolution2019}
\end{figure*}

Finally, Figures~\ref{topic_46_evolution2019}, \ref{topic_63_evolution2015}, \ref{topic_80_evolution2015} and \ref{topic_64_evolution2019} have the aim of validating the political affiliation of the users. These figures show topics whose main content is the support to a specific party, and we observe that the majority of the users active in each topic are the supporters of the corresponding party. However, Fig.~\ref{topic_63_evolution2015} shows how M. Macri, candidate from \textit{Cambiemos} (C), received support from UNA and PR between the first round and the \textit{ballotage} in 2015, which would be decisive for his final triumph. On the contrary, in Fig.~\ref{topic_80_evolution2015} we observe scarce support for D. Scioli, from the other major party.

\subsection{Other similarities for the 2015 election}

\begin{figure*}[ht]
  \centering
  \includegraphics[width=16cm]{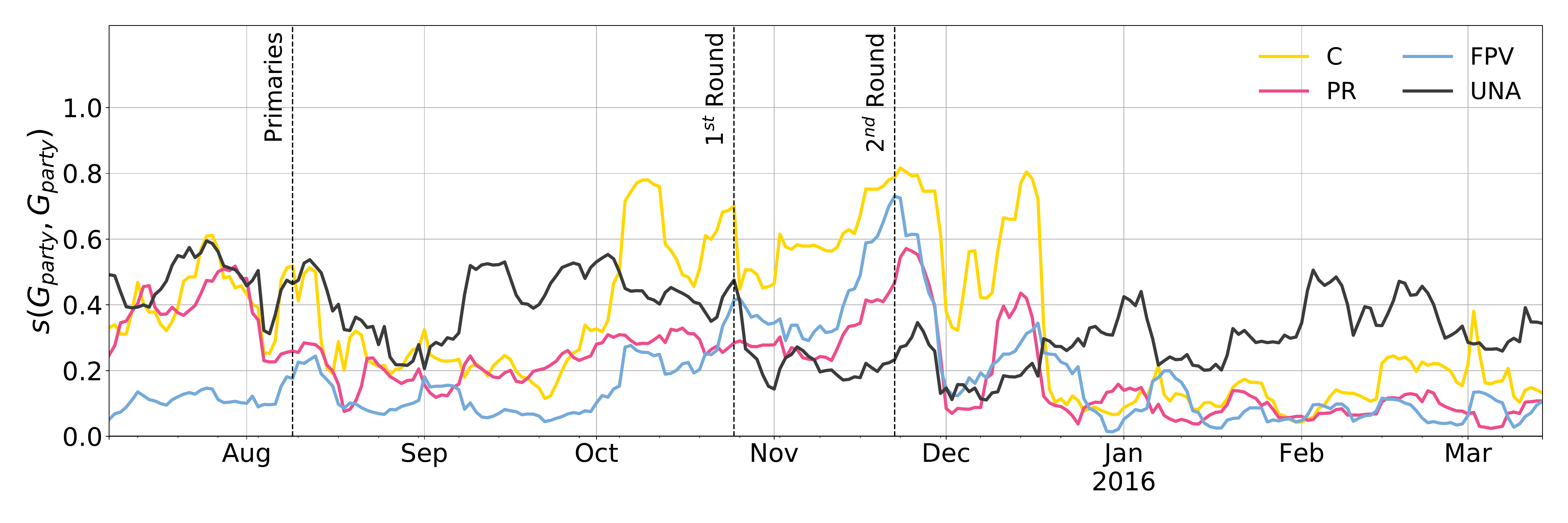}
  \includegraphics[width=16cm]{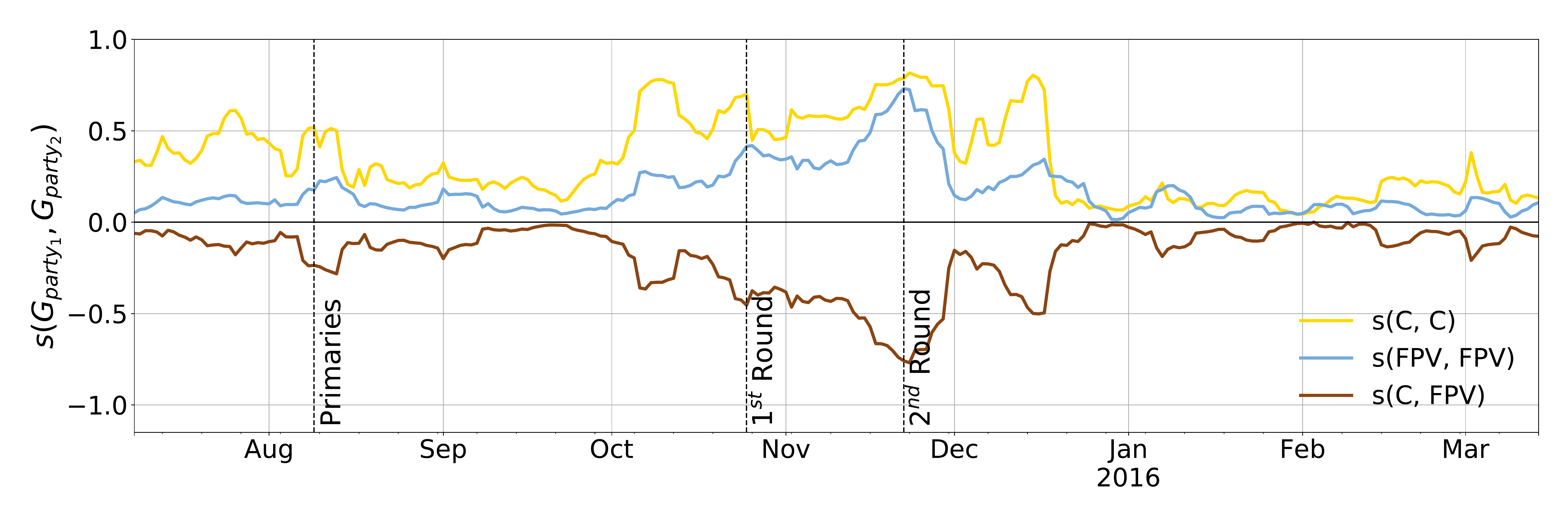}
\caption{\textbf{Similarity among groups of supporters during the 2015 presidential campaign}. \textit{(a)} Self-similarity for each party; \textit{(b)} Cross-similarity between the two main parties.}
\label{2015}
\end{figure*}

In Figure~\ref{2015} we show analogous figures to the ones in the main text for the 2015 election: the self-similarity for each party during the period, and the cross-similarity between the two main parties. The latter reflects the same pattern as in 2019: the cross-similarity between the two main competing parties is consistently negative, reaching its minimum in the ballotage.

\end{document}